\documentclass[prx,superscriptaddress,twocolumn,longbibliography]{revtex4}

\usepackage{amsmath}
\usepackage{amssymb}
\usepackage{graphicx}
\usepackage{multirow}
\usepackage{subfig}
\usepackage{bbm,color,ulem}
\usepackage{dsfont}
\usepackage{float}
\usepackage{braket}
\allowdisplaybreaks

\newcommand{\ua}{\uparrow}
\newcommand{\da}{\downarrow}
\newcommand{\ud}{\upharpoonleft\!\downharpoonright}

\newcommand{\numb}{\addtocounter{equation}{1}\tag{\theequation}}

\begin{document}
\title{Spin-Valley Qubit Dynamics In Exchange Coupled Silicon Quantum Dots}

\author{Donovan Buterakos}
\author{Sankar Das Sarma}
\affiliation{Condensed Matter Theory Center and Joint Quantum Institute, Department of Physics, University of Maryland, College Park, Maryland 20742-4111 USA}
\date{\today}

\begin{abstract}
	The presence of valley states is a significant obstacle to realizing quantum information technologies in Silicon quantum dots, as leakage into alternate valley states can introduce errors into the computation. We use a perturbative analytical approach to study the dynamics of exchange-coupled quantum dots with valley degrees of freedom. We show that if the valley splitting is large and electrons are not properly initialized to valley eigenstates, then time evolution of the system will lead to spin-valley entanglement. Spin-valley entanglement will also occur if the valley splitting is small and electrons are not initialized to the same valley state. Additionally, we show that for small valley splitting, spin-valley entanglement does not affect measurement probabilities of two-qubit systems; however, systems with more qubits will be affected. This means that two-qubit gate fidelities measured in two-qubit systems may miss the effects of valley degrees of freedom. Our work shows how the existence of valleys may adversely affect multiqubit fidelities even when the system temperature is very low.
\end{abstract}

\maketitle

\section{Introduction}

Silicon quantum dots have shown to be a promising candidate system for realizing quantum information technologies due to their long coherence times, fast gate times, potential for scalability, and integration within the current semiconductor industry. Significant progress has been made in the study of Si-based quantum-dot type qubits either using Si-MOS or Si-Ge devices including these representative (but by no means exhaustive) recent experimental publications \cite{HuangNat2019,MaurandNC2016,ZhaoNC2019,YangNat2020,PetitNat2020,FogartyNC2018,VeldhorstNat2015,ZajacSci2018,WatsonNat2018,ZajacPRApp2016,MillsNC2019,BorjansNat2020,SigillitoNPJQI2019,XuePRX2020,KawakamiNN2014,XuePRX2019}. An eventual large-scale quantum computer will solve problems which no classical digital computers can.  One such problem, which ushered in the modern era of quantum computing, is the Shor algorithm for prime factorization \cite{ShorSIAMJC1997}. This is a problem of great technological interest as it is used in all modern cryptography. Trying to factorize a 1000-digit number, which classical computers cannot since the computation cost is exponential in the number of digits, one may need roughly a million logical qubits. Since quantum error correction is essential for quantum computing to actually work, each logical qubit may easily require several thousand physical qubits for its realization, so in the end a hypothetical quantum CPU trying to decisively beat classical computers in doing prime factorization may require $\sim 10^{10}$ physical qubits. Although this sounds like a huge number of qubits, an ordinary CPU chip today may host $10^{10}$ transistors or bits. Each of these transistors or classical bits is made of silicon, giving Si a huge materials and technological advantage in building a quantum computer. In this context, it is understandable why there are multiple large groups all over the world (e.g. INTEL, Princeton, Wisconsin, Sandia, Delft, Sydney, Hefei...) involved in developing Si-based qubits although in terms of the number of working qubits today, the Si system is well behind ion trap or superconducting transmon qubits. So far, only 1-4 Si qubits have been successfully demonstrated in the laboratory.

The physical property being used to create the Si qubit is the electron spin localized within an effective Si quantum dot near a Si surface or interface with another material, such as Si-SiO2 MOS system or Si-Ge 2D electron system.  Since electron spin is by definition a quantum 2-level system, an isolated localized electron spin, if it can be manipulated without much decoherence, is an ideal qubit. Silicon has enormous advantage because electron spins in Si are relatively long-lived with long spin coherence times even for natural Si, which can be  enhanced greatly by isotopic purification \cite{WitzelPRL2010}.  In addition, 2-qubit gates can be implemented by exchange coupling neighboring localized spins by electrically controlling the tunnel coupling between neighboring quantum dots, allowing very fast gate operations.  Thus, long spin coherence time, fast electrostatic gating, and the existing Si chip technology allowing scaling up in principle make Si-based quantum information processing extremely attractive. There is however one serious fundamental problem: Valley. Bulk Si has six equivalent conduction band minima, with the ground state of Si quantum dots having two valleys which are energetically degenerate in the ideal limit.  Thus, ground state Si quantum dot electron spin can in principle be any of the four degenerate spin-valley states.  Typically, there is always some valley splitting associated with the surface/interface \cite{SaraivaPRB2009,SaraivaPRB2011}, but the magnitude of this valley splitting is uncontrollable and it varies randomly from dot to dot in essentially a random manner.  The valley problem in silicon quantum dots, which has received little attention in the literature so far although most researchers in the field recognize its importance \cite{WitzelPRL2010,SaraivaPRB2009,SaraivaPRB2011,CulcerPRL2012,DodsonARXIV2021,CorriganARXIV2020}, is the topic of the current theoretical work.

The Si valley problem has mostly been considered in the context of the valley splitting in the quantum dot qubit being large compared with the qubit temperature ($\sim$ 25-100 mK) so that the thermal occupancy of the higher valley states remains negligible, enabling a valid 2-level quantum description of the system in terms of only the electron spin states.  This is indeed a serious potential problem as the qubit can no longer be defined if higher valley states are occupied.  But, this may not be the only problem when two-qubit gate operations are carried out using inter-qubit exchange coupling. We find that the relative value of the valley splitting with respect to the exchange coupling becomes an important limiting factor even at $T=0$ when any valley splitting is by definition much larger than the temperature. In particular, we find that the valley splitting must be much larger than the inter-qubit exchange coupling to avoid leakage (i.e. quantum decoherence) for the Si system to operate as a multi-qubit quantum computing platform, and we also find that even when the valley splitting is large, the initialization of the valley states becomes a crucial consideration in multiqubit gate operations. This problem of valley splitting in the context of multiqubit gate operations as determined by the exchange coupling between quantum dots has not attracted attention yet in the literature perhaps because of two reasons: (1) There are very few reports of two-qubit exchange gate operations in Si quantum dot qubits; (2) the currently achieved values for the exchange coupling are very small so that the condition of a valley splitting being larger than the exchange energy is automatically satisfied when the valley splitting is larger than temperature.  Since the speed of the 2-qubit gate operations is determined by the exchange coupling strength, higher exchange coupling strength is desirable in the future for progress in Si-based quantum computing, and we want to alert the community that the valley splitting issue is fundamental to 2-qubit gate operations as a totally distinct problem from the one involving thermal occupancies of higher valley states.

The problem with having multiple valley states is that their presence can lead to leakage out of the computational space if the valley degeneracy is not adequately broken. Of the six valley states present in bulk Silicon, four of these decouple when strain is applied to the sample, but the degeneracy of the final two valley states is broken only by a small valley splitting term that is dependent on the microscopics of the system\cite{BorjansPRXQ2021}. We emphasize that there is no known way to control this valley splitting in specific qubits, and in fact, one can figure out the size of the valley splitting only aposteriori.  There is no existing in situ sample diagnostics providing the valley splitting information for the working qubits beforehand.  In the current work, we analyze the effect of valleys on the coupled qubit dynamics, finding that valleys are a much bigger problem for coupled qubits than has been realized so far, and the valley problem worsens radically as the number of qubits goes beyond two.  In fact, we believe that the Si quantum computing community should worry about the valley problem now before building circuits with tens of coupled quantum dots and finding out that they do not work because of the valley problem. The problem of principle we have uncovered here can be `fixed' by having small exchange coupling ($\ll$ valley splitting), but this means that the 2-qubit gate operations will remain bounded by the valley splitting energies.

In this paper, we use a Hubbard model to determine the dynamics of a system of two exchange-coupled quantum dots. We show that spin-valley entanglement can easily arise from time evolution of the system, which is detrimental to the use of the system for quantum information applications. This can be avoided if the valley splitting is large and electrons can be initialized to valley eigenstates, and in fact we show that under these conditions the system maintains coherence and is unaffected to leading order by the valley degree of freedom. If any electron begins in a superposition of valley states then spin-valley entanglement will result, but if the valley splitting is sufficiently large, electrons can be properly initialized to the valley ground state avoiding this situation. However, if the valley splitting is small compared to the exchange interaction strength, then an undesirable spin-valley entanglement is more difficult to avoid, as it will be present unless all electrons are initialized to the same valley state, which is generally difficult to achieve without a large valley splitting.

Additionally, we show that when the valley splitting is small, the measurement probabilities of a two-qubit system are unaffected and are identical to the corresponding measurement probabilities in an ideal system without valley degrees of freedom. However, this is not the case in systems with more than two qubits, and we give examples of gate sequences which give different measurement probabilities in a system with valley states than in an ideal system. Because valley degeneracy affects larger systems but not two-qubit systems, two-qubit gate fidelities measured in two-qubit systems may not accurately account for the effects of valley states, as these effects are only observable in larger system sizes containing more than just two qubits. This last property we discover has not been mentioned in the literature at all, and there has been a feeling that if the 2-qubit gates work, the valley degeneracy problem is irrelevant.  We show that this is false-- one could have perfectly working 2-qubit gates, but the system will lose quantum information through spin-valley entanglement as one scales up to more qubits. This is a very serious issue requiring a resolution before more qubits are added to the circuit. We work at $T=0$ throughout so that the well-understood problem of the thermal occupancy of higher valley states is a non-issue.  We consider the valley degeneracy question only in the context of  gate operations driven by the inter-dot exchange coupling. Just to avoid any misunderstanding, our definition of `large' and `small' valley splitting is as compared with the inter-dot exchange coupling, and not compared with temperature as we are at $T=0$.

This paper is organized as follows: in Sec. II, we present our model and give the Hamiltonian that we use. In Sec. III, we diagonalize the Hamiltonian for a system of two electrons in two quantum dots, for both triplet and singlet spin configurations. In Sec. IV, we discuss the dynamics of the two-qubit system, first for the case where the valley splitting is large, then also for the case of small or zero valley splitting. In Sec. V, we give examples of gate sequences in four-qubit systems where valley effects are observable even when they are not detectable in the 2-qubit situation, and we give our conclusions in Sec. VI.

\section{Model and Hamiltonian}

The Fermi surface of Silicon contains six electron pockets, leading to a six-fold degeneracy in the band structure minima. By applying tensile strain to the sample, two of these valleys are energetically separated from the other four, but remain nearly degenerate to each other. Therefore, in addition to its spin, an electron in a Silicon quantum dot will contain a valley quantum number denoting whether it fills the $\ket{+z}$ or $\ket{-z}$ valley state. Microscopic features of the system introduce a small position-dependent valley splitting. We consider a double quantum dot in Silicon, which yields the following effective Hubbard Hamiltonian \cite{BorjansPRXQ2021,DasSarmaPRB2011,YangPRB2011}:

\begin{align*}
&H=\!\!\sum_{s\in\{\ua,\da\}}\sum_{i=1}^2\Big(\tilde{\Delta}_ic^\dagger_{i,+z,s}c_{i,-z,s}+\tilde{\Delta}_i^*c^\dagger_{i,-z,s}c_{i,+z,s}\Big)\\&+\frac{\epsilon}{2}(n_1-n_2)+t_c\!\!\!\sum_{s\in\{\ua,\da\}}\sum_{v=\pm z}\Big(c^\dagger_{1,v,s}c_{2,v,s}+c^\dagger_{2,v,s}c_{1,v,s}\Big)\\&+\sum_{i=1}^2\frac{U}{2}n_i(n_i-1)
\numb
\end{align*}

where $c_{i,v,s}$ is the second quantized annihilation operator for dot $i$, valley state $v$, and spin $s$, and where  $n_i=n_{i,+z,\ua}+n_{i,+z,\da}+n_{i,-z,\ua}+n_{i,-z,\da}$. Here $\epsilon$ is the detuning between the two quantum dots, $t_c$ is the tunneling constant between the two dots, and $\tilde{\Delta}_i$ determines the valley splitting of dot $i$.  We define $U$ to be the Coulomb energy difference between a dot occupied by two electrons, and a state with one electron occupying each dot. Thus we use a short-range Coulomb interaction term which is independent of valley states (including an explicit valley dependence in the interaction complicates the calculation, but does not affect our conclusion). Let $\tilde{\Delta}_i=\Delta_ie^{-i\phi_i}$, where $\Delta_i=|\tilde{\Delta}_i|$. Define $\ket{i_\pm}=(\ket{i,+z}\pm e^{i\phi_i}\ket{i,-z})/\sqrt{2}$. Here $\ket{i_\pm}$ are the eigenstates of a single electron in a single dot $i$. Then for a single electron in a double-quantum-dot, transforming $H$ into the basis $\{\ket{1_+},\ket{1_-},\ket{2_+},\ket{2_-}\}$ gives the following:

\begin{equation}
H=\begin{pmatrix}
\epsilon/2+\Delta_1&0&t_+&t_-\\
0&\epsilon/2-\Delta_1&t_-&t_+\\
t_+^*&t_-^*&-\epsilon/2+\Delta_2&0\\
t_-^*&t_+^*&0&-\epsilon/2-\Delta_2
\end{pmatrix}
\end{equation}

where $t_\pm=(1\pm e^{i(\phi_2-\phi_1)})t_c/2$. This is equivalent to the two-dot Hamiltonian given in Ref. \onlinecite{BorjansPRXQ2021} up to a constant energy shift.  We emphasize that this Hubbard model Hamiltonian is in fact an accurate description of quantum dot qubit coupling in reasonably realistic situations \cite{DasSarmaPRB2011,YangPRB2011,StaffordPRL1994}.

\section{Diagonalization of Hamiltonian}

In order to determine the effects of valley degeneracy and splitting on the exchange coupling between two dots, we consider two electrons in this two-dot system, diagonalizing the resulting Hamiltonian. We do this separately for the case when the electrons form a spin triplet and when they form a spin singlet.

\subsection{Triplet Spin Configuration}

Consider two electrons with a triplet spin configuration. Then due to the Pauli exclusion principle, the electrons must occupy different orbital or valley states. We will assume that $U\gg t_c,\Delta_j,\epsilon$, as the short-range Coulomb interaction energy is the largest energy scale in the system. Then there are four low energy states: $\ket{1_+2_+}$, $\ket{1_+2_-}$, $\ket{1_-2_+}$, and $\ket{1_-2_-}$. These states couple via tunneling to the two high energy states $\ket{1_+1_-}$ and $\ket{2_+2_-}$, where the electrons occupy both valley states in a single dot. Perturbation theory to first order in $U^{-1}$ gives the following Hamiltonian for the four lowest-energy states after including the effects of coupling to the two high energy states:

\begin{widetext}
\begin{equation}
H_T=\begin{pmatrix}
\Delta_1+\Delta_2-\frac{2|t_-|^2}{U}&\frac{t_-t_+^*-t_+t_-^*}{U}&\frac{t_+t_-^*-t_-t_+^*}{U}&\frac{2|t_-|^2}{U}\\
\frac{t_+t_-^*-t_-t_+^*}{U}&\Delta_1-\Delta_2-\frac{2|t_+|^2}{U}&\frac{2|t_+|^2}{U}&\frac{t_-t_+^*-t_+t_-^*}{U}\\
\frac{t_-t_+^*-t_+t_-^*}{U}&\frac{2|t_+|^2}{U}&-\Delta_1+\Delta_2-\frac{2|t_+|^2}{U}&\frac{t_+t_-^*-t_-t_+^*}{U}\\
\frac{2|t_-|^2}{U}&\frac{t_+t_-^*-t_-t_+^*}{U}&\frac{t_-t_+^*-t_+t_-^*}{U}&-\Delta_1-\Delta_2-\frac{2|t_-|^2}{U}
\end{pmatrix}
\label{eqn:ht4}
\end{equation}
\end{widetext}

Note that to leading order in $\epsilon/U$, the four low energy states are independent of $\epsilon$. This is because $\epsilon$ only affects the energies of states where both electrons occupy the same dot, and these are already energetically separated from the other states by the large onsite Coulomb interaction strength $U$ and hence do not contribute in the leading order. If $\epsilon$ is allowed to be of the same scale as $U$, as is done in some experiments to control the exchange interaction, then these results must be adjusted accordingly. Specifically where $U$ appears in the denominator of terms in Eq. (\ref{eqn:ht4}) it must be replaced with $U\pm\epsilon$ depending on the which state introduces each term. However for the purpose of this paper we will focus on the situation where $\epsilon\ll U$, as is the case in experiments which use barrier control of the exchange interaction. In order to further study the system dynamics, Eq. (\ref{eqn:ht4}) must be fully diagonalized. Since this cannot be easily analytically done for arbitrary $\Delta_i$ and $t_c$, we instead consider two different limits for small and large $\Delta_i$. In the limit where $\Delta_i\gg t_c^2/U$, the matrix in eq. (\ref{eqn:ht4}) is already diagonalized to leading order in $t_c^2/U\Delta_i$, and the energies are given by its diagonal entries. In the limit where $\Delta_i\ll t_c^2/U$, diagonalizing Eq. (\ref{eqn:ht4}) yields the following energies:

\begin{align*}
&E_1=-\frac{4t_c^2}{U}\\
&E_2=-|\tilde{\Delta}_1+\tilde{\Delta}_2|\\
&E_3=0\\
&E_4=|\tilde{\Delta}_1+\tilde{\Delta}_2|=\sqrt{\Delta_1^2+\Delta_2^2+2\Delta_1\Delta_2\cos(\phi_2-\phi_1)}
\numb
\label{eqn:et}
\end{align*}

The corresponding eigenstates are given by:

\begin{widetext}
\begin{align*}
&\ket{\psi_1}=\frac{1}{\sqrt{2}t_c}\Big(t_-\ket{1_+2_+}-t_+\ket{1_+2_-}+t_+\ket{1_-2_+}-t_-\ket{1_-2_-}\Big)\\
&\ket{\psi_2}=\Big[\frac{(\Delta_2+\Delta_1)t_+}{2t_c|\tilde{\Delta}_1+\tilde{\Delta}_2|}\big(\ket{1_+2_+}+\ket{1_-2_-}\big)+\frac{(\Delta_2-\Delta_1)t_-}{2t_c|\tilde{\Delta}_1+\tilde{\Delta}_2|}\big(\ket{1_+2_-}+\ket{1_-2_+}\big)\\&\qquad\qquad-\frac{1}{2t_c}\Big(t_+\ket{1_+2_+}-t_-\ket{1_+2_-}+t_-\ket{1_-2_+}-t_+\ket{1_-2_-}\Big)\Big]\\
&\ket{\psi_3}=\frac{1}{\sqrt{2}t_c|\tilde{\Delta}_1+\tilde{\Delta}_2|}\Big[(\Delta_2-\Delta_1)t_-\big(\ket{1_+2_+}+\ket{1_-2_-}\big)+(\Delta_2+\Delta_1)t_+\big(\ket{1_+2_-}+\ket{1_-2_+}\big)\Big]\\
&\ket{\psi_4}=\Big[\frac{(\Delta_2+\Delta_1)t_+}{2t_c|\tilde{\Delta}_1+\tilde{\Delta}_2|}\big(\ket{1_+2_+}+\ket{1_-2_-}\big)+\frac{(\Delta_2-\Delta_1)t_-}{2t_c|\tilde{\Delta}_1+\tilde{\Delta}_2|}\big(\ket{1_+2_-}+\ket{1_-2_+}\big)\\&\qquad\qquad+\frac{1}{2t_c}\Big(t_+\ket{1_+2_+}-t_-\ket{1_+2_-}+t_-\ket{1_-2_+}-t_+\ket{1_-2_-}\Big)\Big]
\numb
\label{eqn:eigenstates}
\end{align*}
\end{widetext}

\subsection{Singlet Spin Configuration}

We now repeat the same calculation for a pair of electrons in a singlet spin configuration. In this case, the same six orbital states as in the triplet case are present, with the addition of four doubly occupied states $\ket{1_{+\ud}}$, $\ket{1_{-\ud}}$, $\ket{2_{+\ud}}$, and $\ket{2_{-\ud}}$, since both electrons can occupy the same valley and orbital state. Like above, we apply perturbation theory in $U^{-1}$, and calculate the Hamiltonian for the four lowest energy states, yielding:

\begin{widetext}
\begin{equation}
H_S=\begin{pmatrix}
\Delta_1+\Delta_2-\frac{2|t_-|^2+4|t_+|^2}{U}&-\frac{2t_-t_+^*}{U}&-\frac{2t_+t_-^*}{U}&-\frac{2|t_-|^2}{U}\\
-\frac{2t_+t_-^*}{U}&\Delta_1-\Delta_2-\frac{2|t_+|^2+4|t_-|^2}{U}&-\frac{2|t_+|^2}{U}&-\frac{2t_-t_+^*}{U}\\
-\frac{2t_-t_+^*}{U}&-\frac{2|t_+|^2}{U}&-\Delta_1+\Delta_2-\frac{2|t_+|^2+4|t_-|^2}{U}&-\frac{2t_+t_-^*}{U}\\
-\frac{2|t_-|^2}{U}&-\frac{2t_+t_-^*}{U}&-\frac{2t_-t_+^*}{U}&-\Delta_1-\Delta_2-\frac{2|t_-|^2+4|t_+|^2}{U}
\end{pmatrix}
\label{eqn:hs4}
\end{equation}
\end{widetext}

Diagonalizing this matrix in the limit where $\Delta_i\ll t_c^2/U$ gives the following energies:

\begin{align*}
&E_1=0\\
&E_2=-\frac{4t_c^2}{U}-|\tilde{\Delta}_1+\tilde{\Delta}_2|\\
&E_3=-\frac{4t_c^2}{U}\\
&E_4=-\frac{4t_c^2}{U}+|\tilde{\Delta}_1+\tilde{\Delta}_2|
\numb
\label{eqn:es}
\end{align*}

The corresponding singlet case eigenstates are precisely the same as for the triplet case given by Eq. (\ref{eqn:eigenstates}). Note that $E_2$, $E_3$, and $E_4$ are less than the corresponding triplet energies; however, $E_1$ is greater than the corresponding triplet energy. In each case, the difference is $\pm J_0$, where $J_0=4t_c^2/U$ is the strength of the exchange interaction in an ideal system which does not have any valley degeneracy.

\section{System Dynamics}

We now investigate the dynamics of a system prepared in a specific initial state and allowed to evolve under the Hamiltonian for some time $t$. This is the coupled qubit dynamics under gate operation, which controls quantum computation. We show that when the initial state is not prepared with each electron in the same valley state, leakage between valley states will introduce error. If the valley splitting $\Delta_i$ is small, this error can occur even if all electrons are initialized in the valley ground states $\ket{i_-}$ if the phases $\phi_i$ differ from one another. We first examine the large valley-splitting limit where $\Delta_i\gg J_0$, followed by the small valley-splitting limit where $\Delta_i\ll J_0$.

\subsection{Large Valley Splitting}

When $\Delta_i\gg t_c^2/U$, the system dynamics is determined to leading order by the diagonal entries of Eqs. (\ref{eqn:ht4}) \& (\ref{eqn:hs4}). The off-diagonal elements only affect the energies to order $J_0^2/\Delta_i$. If the system is prepared in one of the valley eigenstates, there will be an effective exchange interaction $J_+=4|t_+|^2/U$ for valley states $\ket{1_+2_+}$ and $\ket{1_-2_-}$; and an effective exchange interaction $J_-=4|t_-|^2/U$ for valley states $\ket{1_+2_-}$ and $\ket{1_-2_+}$. Thus to leading order in $J_0/\Delta_i$, the presence of energetically separated valley states does not affect the dynamics of the system as long as the initial state is a valley eigenstate.

If not all electrons are initialized to valley eigenstates, then the time evolution of the system can give rise to spin-valley entanglement. Consider, for example, an initial state where qubit 1 starts in the $\ket{1_{+\ua}}$ state, but qubit 2 starts in the state $(\ket{2_{+\da}}+\ket{2_{-\da}})/\sqrt{2}$. Then after some time $t$, the system will evolve to:

\begin{align*}
\ket{\Psi(t)}&=\frac{1}{2}\ket{1_+2_+}\Big(\ket{T_0}+e^{iJ_+t}\ket{S}\Big)\\&+\frac{1}{2}e^{it(2\Delta_2+\frac{J_+-J_-}{2})}\ket{1_+2_-}\Big(\ket{T_0}+e^{iJ_-t}\ket{S}\Big)
\numb
\label{eqn:timeevv2}
\end{align*}

where $\ket{T_0}$ and $\ket{S}$ are the triplet and singlet spin states respectively. Suppose for simplicity that $\phi_1=\phi_2$, and thus $J_-=0$ and $J_+=J_0$. Then the system reaches a maximally entangled state when $J_0t=(2k+1)\pi$ for integer $k$:

\begin{align*}
\ket{\Psi\big(\frac{(2k+1)\pi}{J_0}\big)}&=\frac{1}{\sqrt{2}}\ket{1_+2_+}\ket{\da\ua}\\&+\frac{1}{\sqrt{2}}e^{it(2\Delta_2+\frac{J_+}{2})}\ket{1_+2_-}\ket{\ua\da}
\numb
\end{align*}

Because the electron in dot 1 was initialized in a valley eigenstate in this example, it remains in that state throughout the evolution of the system. Thus, the state at any particular point in time corresponds a point on each of two Bloch spheres which represent the combined spin state (singlet or triplet) and the valley state of the second electron. This is not a one-to-one correspondence, as different entangled states can correspond to the same set of points; nevertheless, it is useful for visualizing the information stored in the spin and valley states and the entanglement between them.

\begin{figure}[!htb]
	\includegraphics[width=.4\columnwidth]{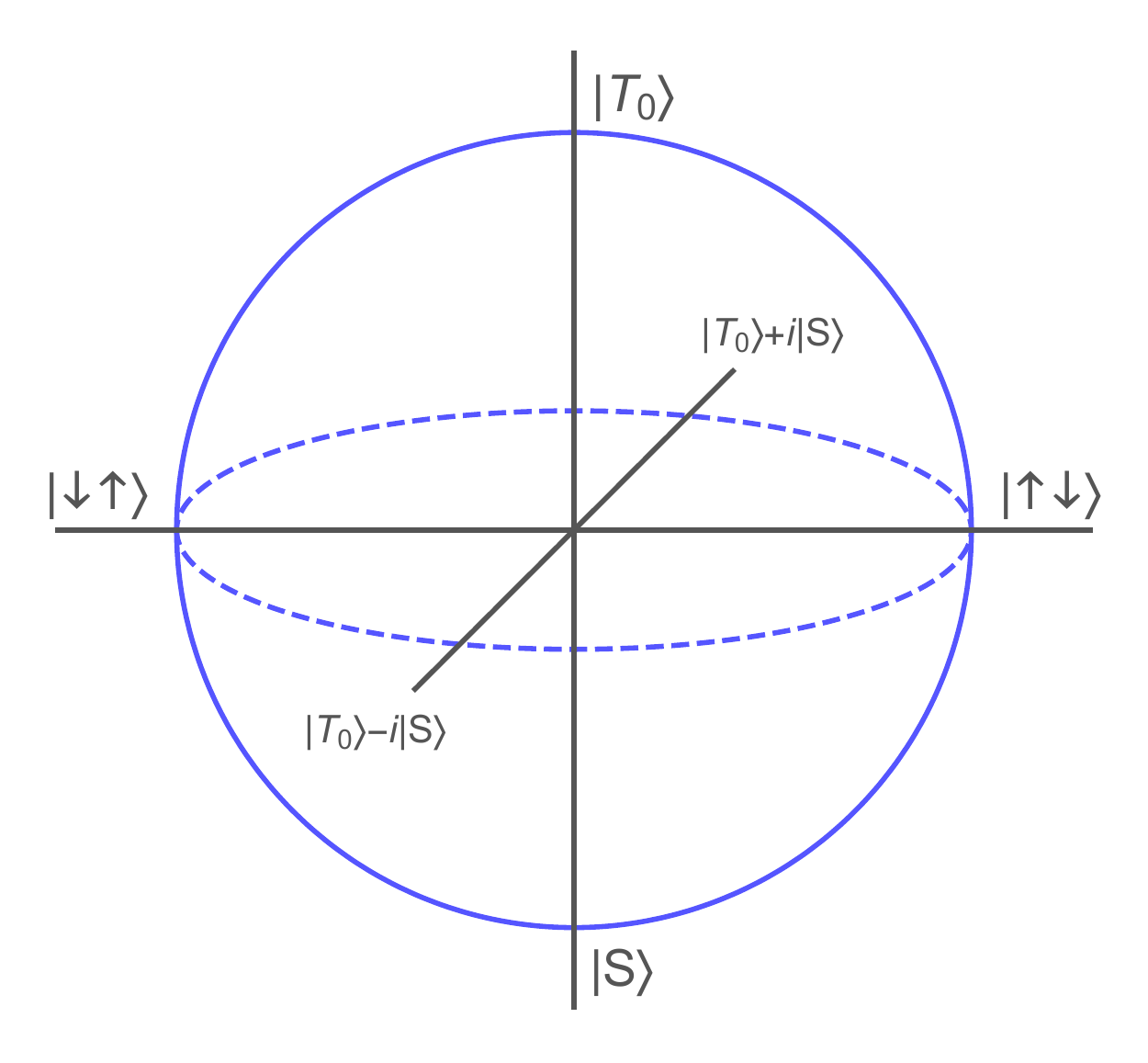}
	\includegraphics[width=.43\columnwidth]{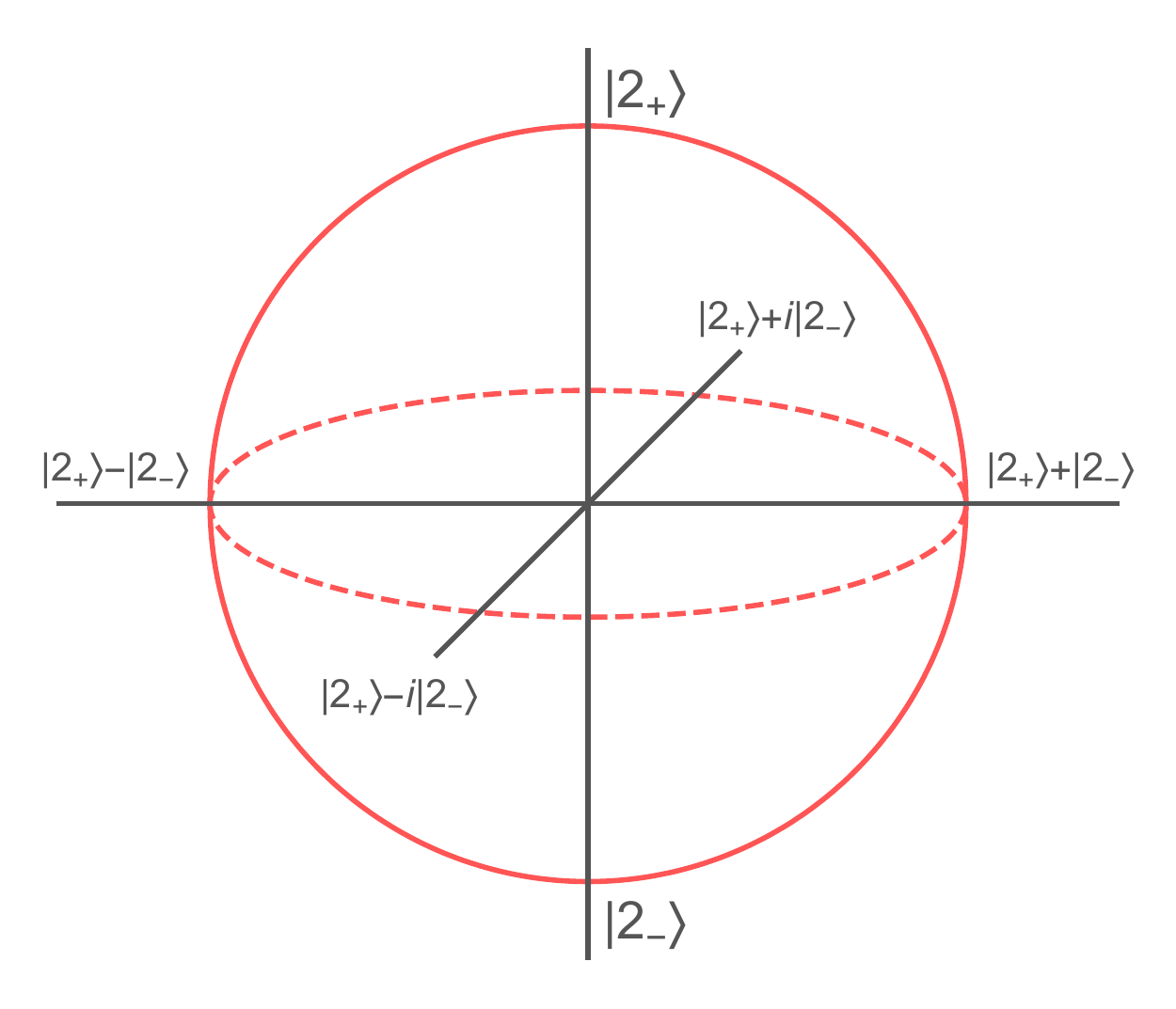}
	\includegraphics[width=.4\columnwidth]{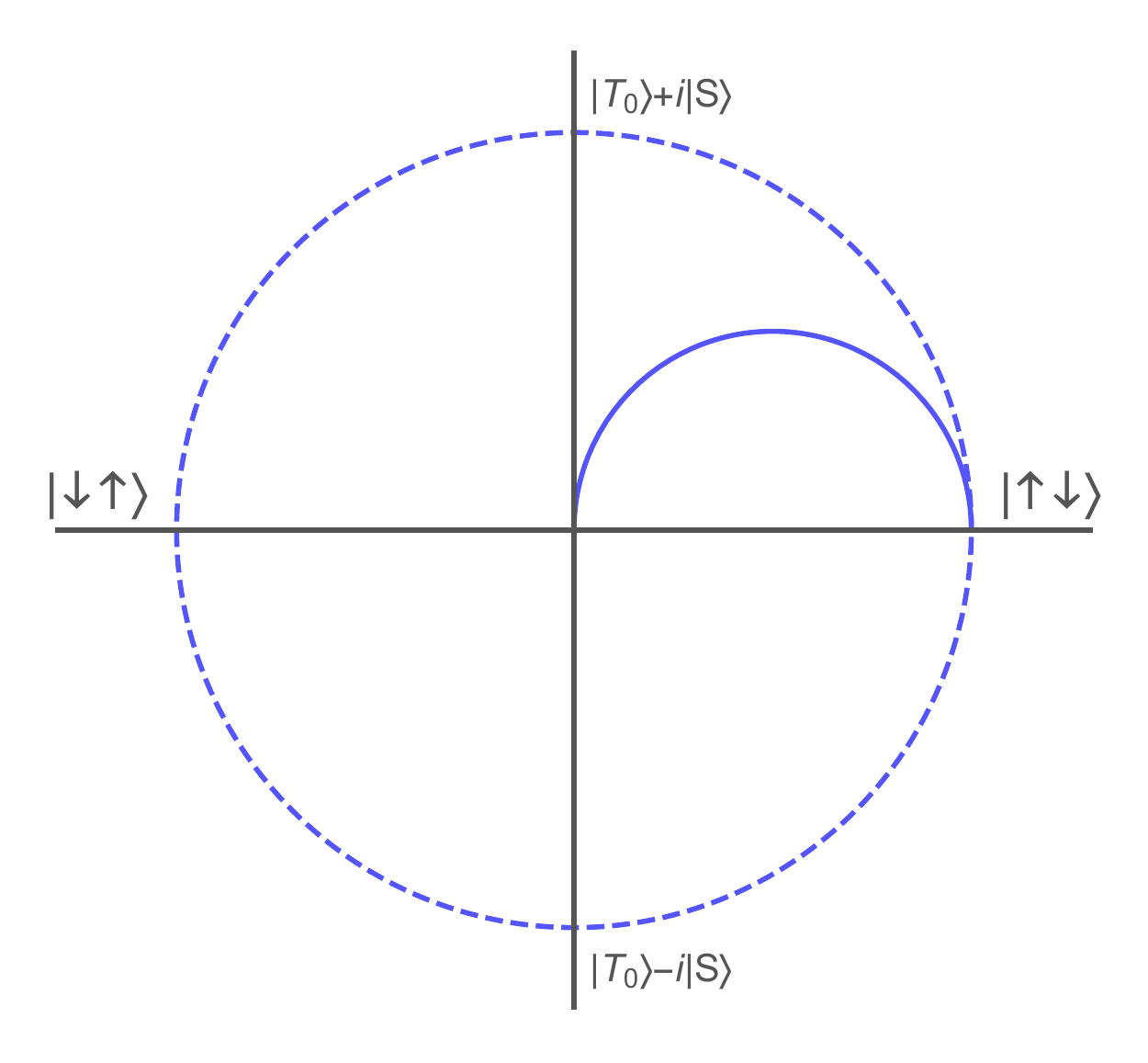}
	\includegraphics[width=.43\columnwidth]{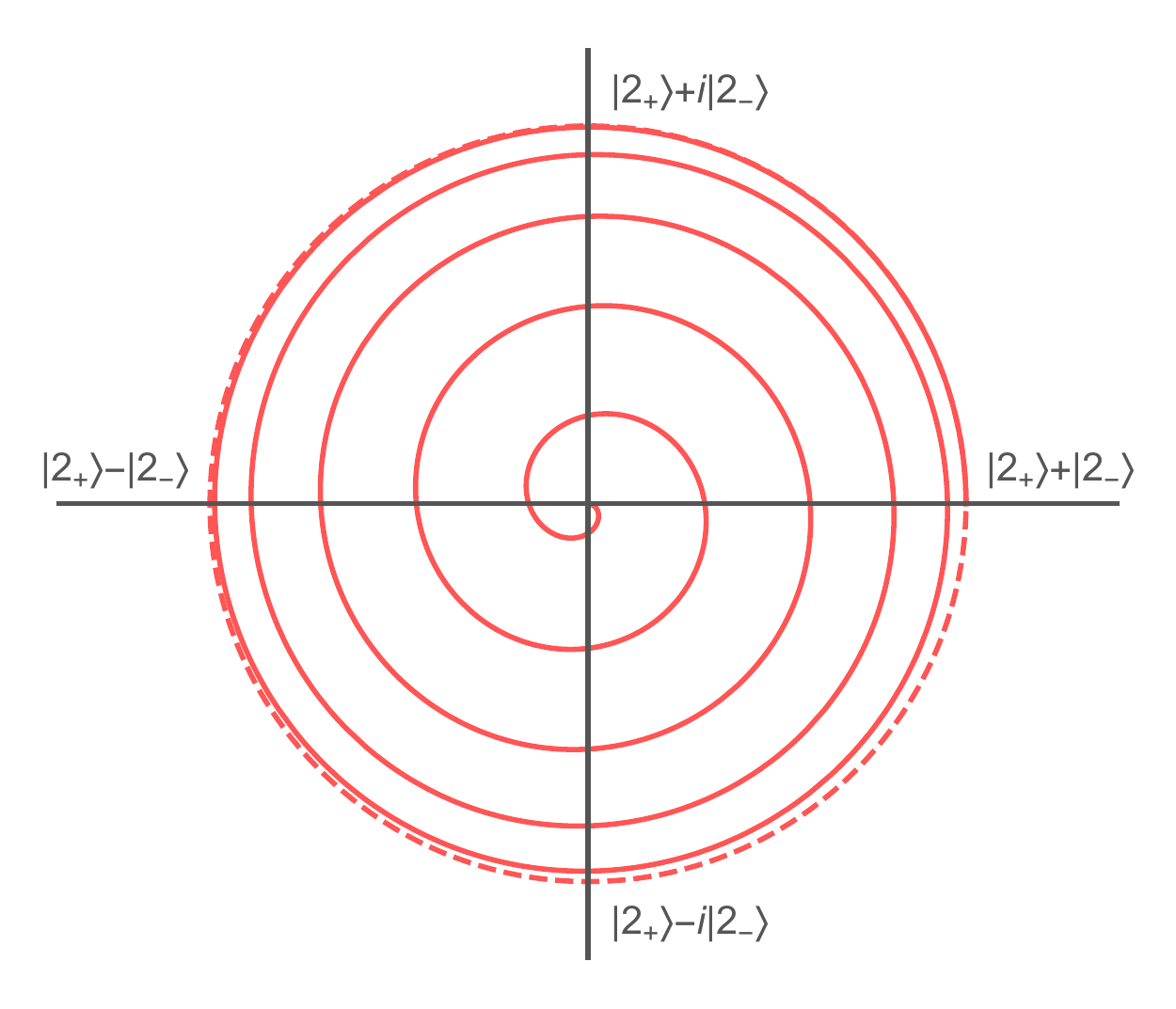}
	\caption{{\bf Top:} Bloch sphere representation of the spin qubit in the singlet-triplet basis and valley qubit of the second dot (the first dot will always be in the valley state $\ket{1_+}$). {\bf Bottom:} Time evolution of a state given by eq. (\ref{eqn:timeevv2}) for $t$ ranging from $0$ to $\pi/J_0$, with $\Delta_2=5J_0$ and $\phi_1=\phi_2$. The $xy$ cross-sections of the Bloch spheres are plotted since the state stays entirely within the $xy$ planes.}
	\label{fig:blochdelta}
\end{figure}

In Fig. \ref{fig:blochdelta}, we plot the path $\ket{\Psi(t)}$ traces on the two Bloch spheres. The path begins on the surface of both Bloch spheres, indicating that initial state is separable. As time evolves to $t=\pi/J_0$, the path spirals toward the center of both Bloch spheres, indicating maximal spin-valley entanglement. In general, spin-valley entanglement is detrimental to quantum information applications, since measuring the spin state without the ability to measure the corresponding valley state can result in a complete loss of qubit information. Thus it is imperative to initialize the system in valley eigenstates -- otherwise, information will leak out unwittingly through spin-valley entanglement during the coupled qubit dynamical evolution.

\subsection{Small Valley Splitting}

We now consider a situation where the valley splitting is small compared to the exchange interaction strength. In this regime the dynamics is dictated by the states and energies in Eqs. (\ref{eqn:ht4}), (\ref{eqn:eigenstates}), and (\ref{eqn:hs4}). Because the singlet and triplet spin configurations share the same valley eigenstates, there is an effective exchange interaction $\pm J_0$ depending on the specific valley states occupied. We show that unless the system is initialized with all electrons occupying the same valley state, that the time evolution of the coupled system will result in spin-valley entanglement. Initializing the system in this way is difficult due to the near-degeneracy of the valley states. Additionally, if the phase difference $\phi_2-\phi_1$ between dots is nonzero, as is often the case,  spin-valley entanglement will occur even if all electrons are initialized to their individual ground valley states. In fact, we are not aware of any experimental control capable of tuning the inter-valley phase difference $\phi_2-\phi_1$, which we see as a potential problem requiring a resolution for Si spin qubits to work in a large circuit with many operational qubits.

To demonstrate the presence of spin-valley entanglement which can arise, consider an initial state $\ket{1_{+\ua}2_{+\da}}$. For simplicity, suppose $\Delta_1=\Delta_2=\Delta$, and suppose there is a nonzero phase difference $\phi_2-\phi_1$. Then after some time $t$, the state will evolve to:

\begin{align*}
\ket{\Psi(t)}&=\frac{t_-^*}{2t_c}\ket{\psi_1}\Big(e^{iJ_0t}\ket{T_0}+\ket{S}\Big)\\&+\frac{t_+^*}{2\sqrt{2}t_c}\bigg[\Big(\frac{t_c}{|t_+|}-1\Big)e^{i\frac{2|t_+|}{t_c}\Delta t}\ket{\psi_2}\\&+\Big(\frac{t_c}{|t_+|}+1\Big)e^{-i\frac{2|t_+|}{t_c}\Delta t}\ket{\psi_4}\bigg]\Big(\ket{T_0}+e^{iJ_0t}\ket{S}\Big)
\numb
\label{eqn:timeevd}
\end{align*}

\begin{figure}[!htb]
	\includegraphics[width=.48\columnwidth]{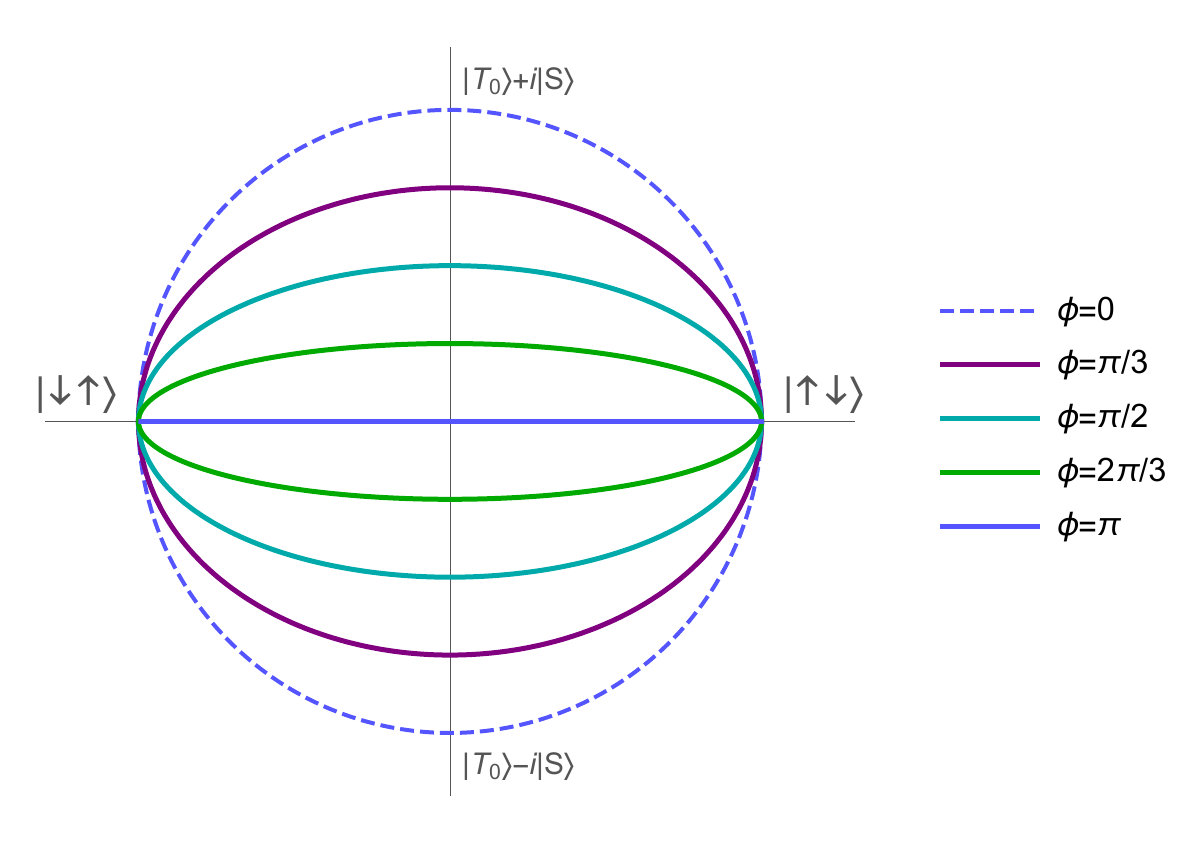}
	\includegraphics[width=.48\columnwidth]{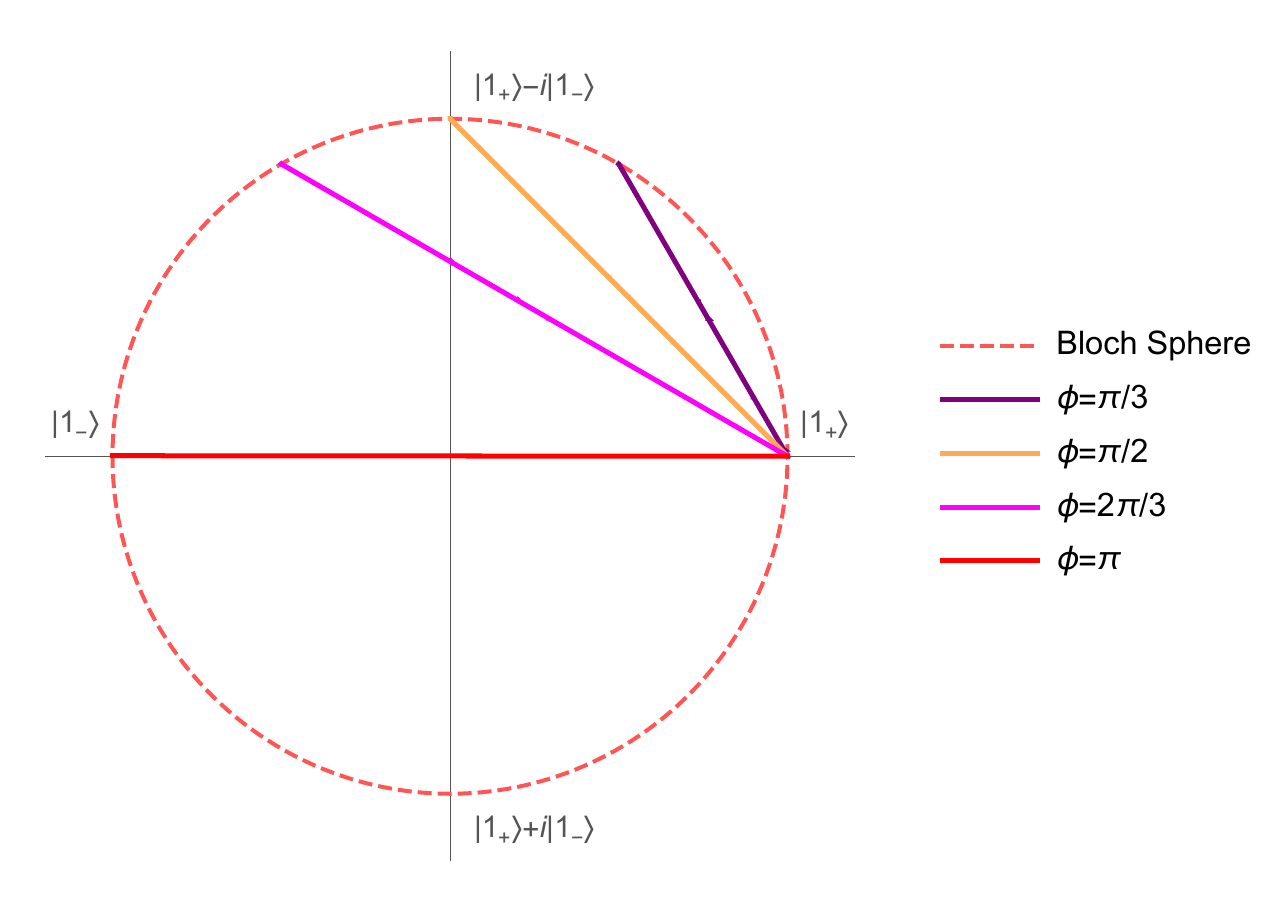}
	\caption{{\bf Left:} Time evolution of the spin state given by Eq. (\ref{eqn:timeevd}) for $t$ ranging from $0$ to $2\pi/J_0$ plotted for different values of $\phi$. The $xy$ cross-section of the Bloch sphere is shown since the path stays entirely within the $xy$ planes. These paths are independent of $\Delta$. {\bf Right:} Projection into the $xy$-plane of the valley state of dot 1 given by Eq. (\ref{eqn:timeevd}), plotted for different values of $\phi$. The projections of these paths are independent of $\Delta$; however, the paths themselves do vary with $\Delta$ as shown in fig. \ref{fig:v1curvd}. The plot for the valley state of dot 2 is identical, but reflected vertically.}
	\label{fig:pathd}
\end{figure}

\begin{figure}[!htb]
	\centering
	\includegraphics[width=.49\columnwidth]{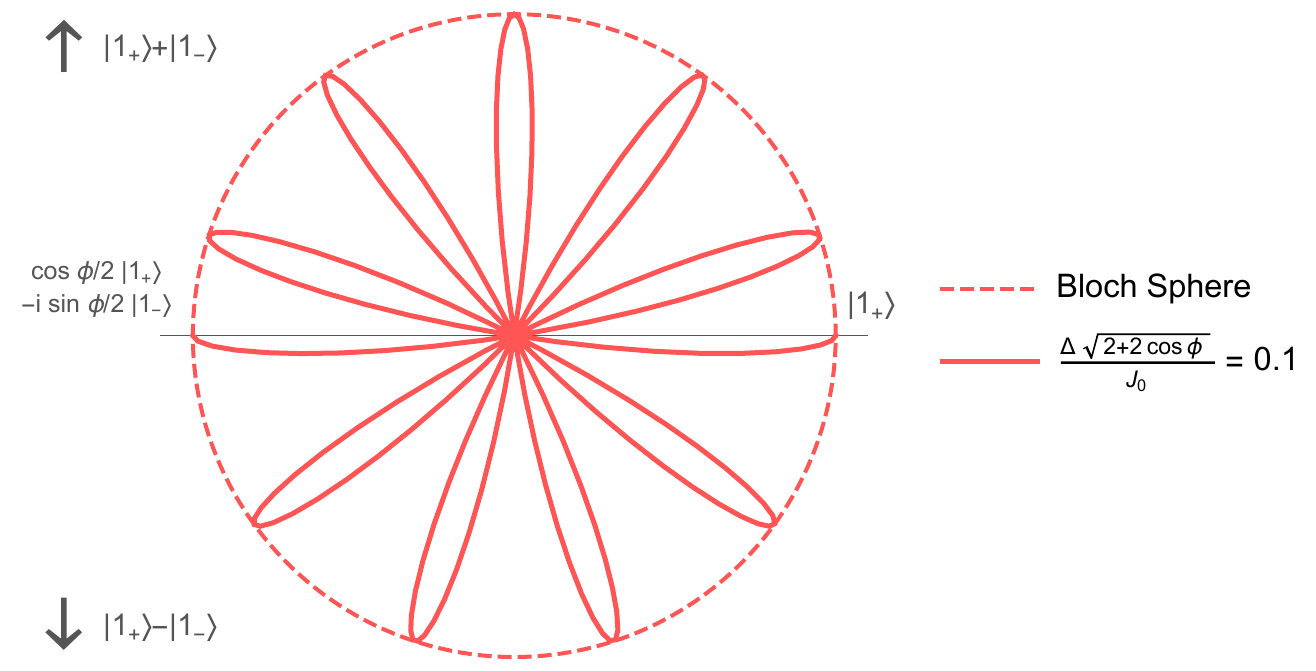}
	\includegraphics[width=.49\columnwidth]{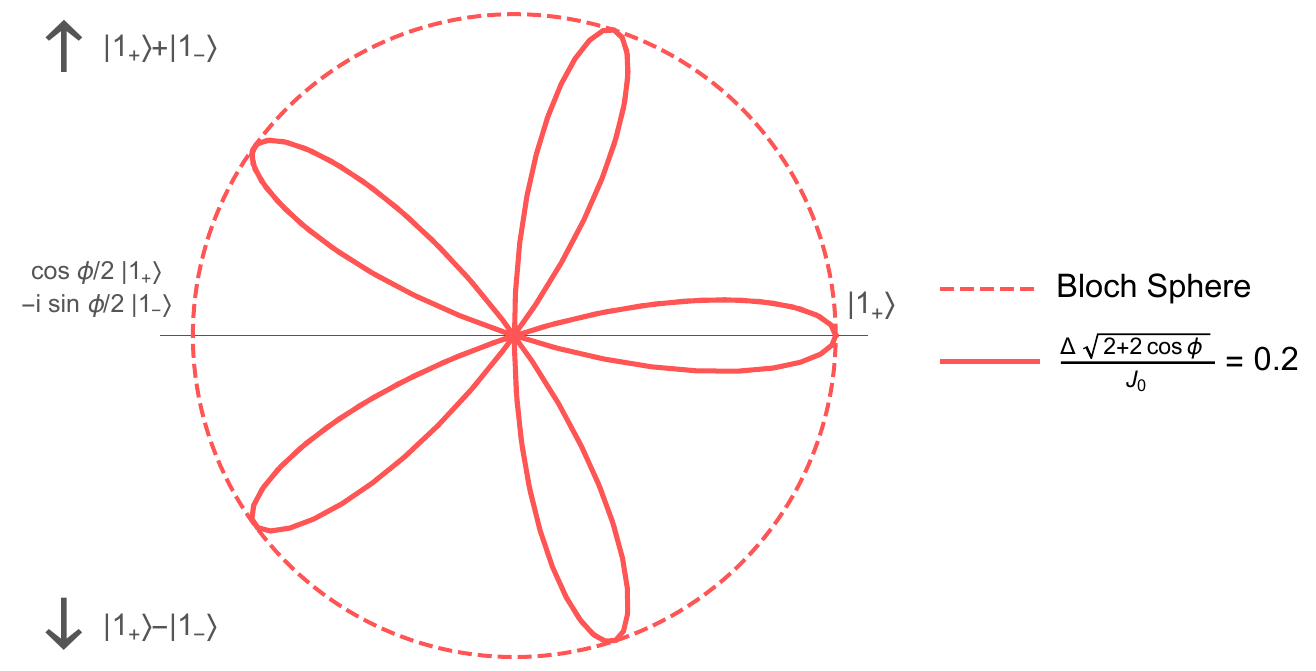}
	\includegraphics[width=.49\columnwidth]{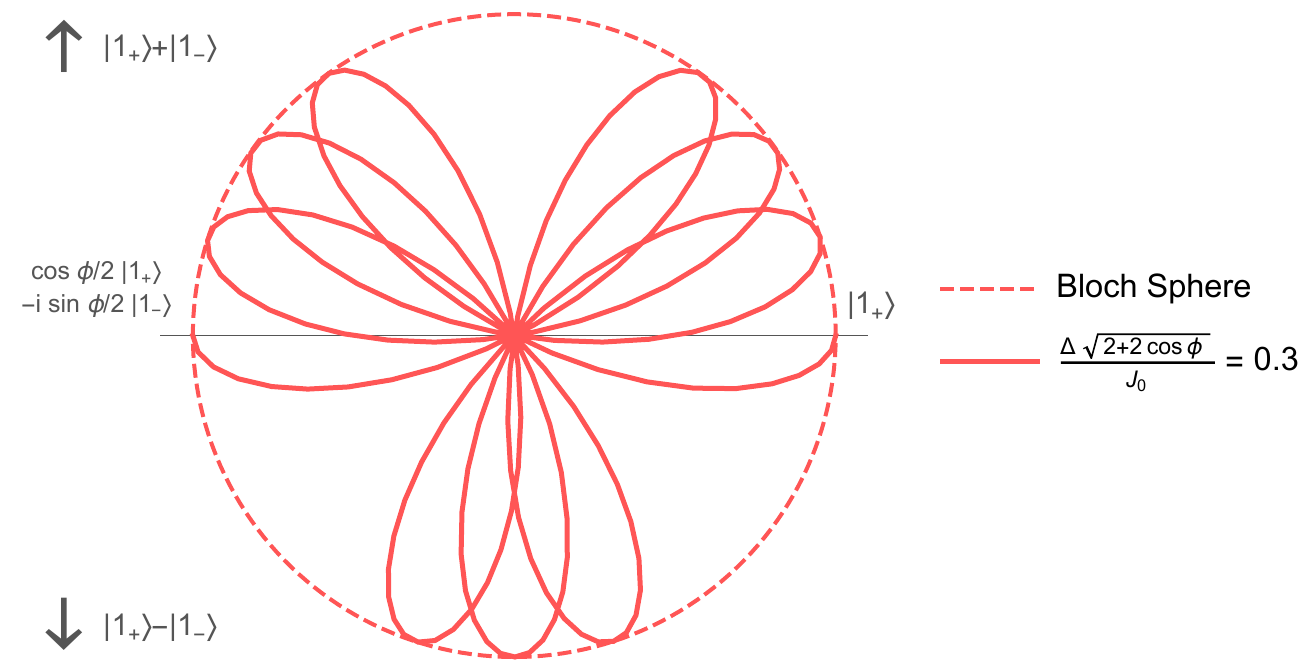}
	\caption{Time evolution of the valley state of dot 1 given by eq. (\ref{eqn:timeevd}). The path which represents this state's time evolution lies entirely within the plane given by $x+y\tan\phi=1$. In these plots, the vertical direction corresponds to the $\pm z$-direction, and the horizontal axis is the line given by $\{x+y\tan\phi=1,z=0\}$, which is the set of lines plotted in Fig. \ref{fig:pathd}. The dashed circle represents the intersection of the Bloch sphere with the plane $x+y\tan\phi=1$ and is not necessarily unit length. The shape of the path depends only on the quantity $\Delta\sqrt{2+2\cos\phi}/J_0$, and we give plots where this quantity equals 0.1, 0.2, and 0.3.}
	\label{fig:v1curvd}
\end{figure}

By definition, $2|t_+|/t_c=\sqrt{2+2\cos(\phi_2-\phi_1)}$, which equals $|\tilde{\Delta}_1+\tilde{\Delta}_2|/\Delta$ when $\Delta_1=\Delta_2$. In Fig. \ref{fig:pathd} we plot the path on the Bloch sphere drawn out by total spin of $\ket{\Psi(t)}$. This path is independent of the value of $\Delta$, and is confined to the $xy$-plane. The path given by the valley states is not confined to the $xy$-plane, but its projection into the $xy$-plane is a chord of the circle, the angle of which is determined by $\phi$. In Fig. \ref{fig:v1curvd}, we plot the full shape of the path, which lies in the plane defined by the specific $\phi$-dependent chord extended in the $z$-direction. 

In general, $\ket{\Psi(t)}$ is an entangled state with two exceptions. $\ket{\Psi(t)}$ is separable if $\phi_1=\phi_2$, as this makes $t_-=0$ causing the first term in Eq. (\ref{eqn:timeevd}) to vanish. $\ket{\Psi(t)}$ also becomes separable when $t=k\pi/J_0$ (with integer k), as this causes the spin states in both terms of Eq. (\ref{eqn:timeevd}) to become identical. Conversely, $\ket{\Psi(t)}$ becomes a maximally entangled state when $\phi_2-\phi_1=\pi$ and $J_0t=(2k+1)\pi/2$, which causes Eq. (\ref{eqn:timeevd}) to simplify to the following:

\begin{align*}
&\frac{1}{2\sqrt{2}}\Big(\ket{1_+2_+}-\ket{1_-2_-}\Big)\Big(i^{2k+1}\ket{T_0}+\ket{S}\Big)\\+\;&\frac{1}{2\sqrt{2}}\Big(\ket{1_+2_+}+\ket{1_-2_-}\Big)\Big(\ket{T_0}+i^{2k+1}\ket{S}\Big)
\numb
\label{eqn:timeevsimp}
\end{align*}

It is instructive to consider the probability of recovering the initial spin state $\ket{\ua\da}$ when measuring the spin of $\ket{\Psi(t)}$ after some time (leaving the valley state unmeasured). From Eq. (\ref{eqn:timeevd}), this probability is found to be given by:

\begin{equation}
P\big(\ket{1_{+\ua}2_{+\da}}\rightarrow\ket{\ua\da}\big)=\frac{1}{2}\big(1+\cos J_0t\big)
\end{equation}

This is precisely the same measurement probability as would result from time evolution in an ``ideal'' one-valley system. Differences occur when measuring in a different basis; for example the probability of obtaining the state $(\ket{\ua\da}+i\ket{\da\ua})/\sqrt{2}$ is given by:

\begin{equation}
P\Big(\ket{1_{+\ua}2_{+\da}}\rightarrow\frac{\ket{\ua\da}+i\ket{\da\ua}}{\sqrt{2}}\Big)=\frac{1}{2}-\frac{1}{2}\Big(1-\frac{|t_-^2|}{t_c^2}\Big)\sin J_0t
\label{eqn:py}
\end{equation} 

If $\phi_1=\phi_2$, meaning the two electrons start in the same valley states, then Eq. (\ref{eqn:py}) reduces to $(1-\sin J_0t)/2$, which is the same result as would be given by an ideal system. However if $\phi_2-\phi_1=\pi$, then the two electrons start in opposite valley states, and Eq. (\ref{eqn:py}) reduces to $P=1/2$, independent of time. In this case when $J_0t=(2k+1)\pi/2$, the probability outcome of measuring the spin in any basis gives $1/2$, as the spin state is maximally entangled with the valley state.

Despite the difference between Eq. (\ref{eqn:py}) and the ideal case, spin-valley entanglement is difficult to observe in a system of two quantum dots. This is because it is difficult to directly measure in the basis containing the state $(\ket{\ua\da}+i\ket{\da\ua})/\sqrt{2}$. Generally, if a quantum algorithm would require such a measurement, the measurement would be performed by applying a $\sqrt{SWAP}$ gate and then measuring in the $z$ basis, which in an ideal system would produce the same result. However, $\sqrt{SWAP}$ gates are performed via the exchange interaction, which can disentangle a spin-valley entangled state. In fact, for two qubits, as long as states are initialized and measured in the z-basis, valley splitting will not affect the measurement probability to first order in $\Delta/J_0$. This is because the eigenstates given by Eq. (\ref{eqn:eigenstates}) simultaneously diagonalize both the triplet and singlet Hamiltonians, Eqs. (\ref{eqn:ht4}) and (\ref{eqn:hs4}). Additionally, the energy difference between the corresponding triplet and singlet energies, Eqs. (\ref{eqn:et}) and (\ref{eqn:es}), is $\pm J_0$ for every eigenstate. Simply measuring in the z-basis cannot distinguish between rotations by $+J_0$ or $-J_0$.

One possible way to make such a distinction between rotations is the following. Start with a state $\ket{\psi_1}$, and perform a partial rotation (perhaps $2\pi/3$) via the exchange interaction $J_0$. Then allow $\ket{\psi_1}$ through some method to be transformed to any of the other states $\ket{\psi_2}$ through $\ket{\psi_4}$. One way in which this might happen is to let the system precess under the valley splitting, but with no exchange interaction present (note that in this case $\Delta\ll J_0$ does not hold). Finally, complete the initial rotation from the first step. In an ideal one-valley case, the system will have undergone one complete rotation. However, in a system with valley splitting, the last part of the rotation will be in the opposite direction as the first part, and thus will not form a complete rotation. This will affect the measurement probabilities in the z-basis. However, it is not possible to complete the second step (rotating $\ket{\psi_1}$ into $\ket{\psi_2}$) in a system of two qubits while keeping $\Delta\ll J_0$ without adding additional terms to the Hamiltonian.

\section{Observing Valley Effects in 4-Qubit Systems}

In the previous discussion, we demonstrated that despite the spin-valley entanglement that occurs, the measurement probabilities in the $\{\ua,\da\}$ basis would be unaffected by the presence of valley states in a two-dot system. However, we now show that in a 4-dot system this is no longer the case. We do this by giving two examples of sequences of operations which will result in a different measurement probability in the presence of valley states than the same operations would in an ideal one-valley system.

Consider the fully degenerate case where $\Delta_1=\Delta_2=0$ (exactly the same sequences apply when $\Delta_i$ are nonzero, but we consider the degenerate case for the sake of simplicity). Time evolution of the exchange interaction can be used to perform $\sqrt{SWAP}$ gates. Consider an array of four quantum dots prepared in the initial state $\ket{\ua_-\da_+\ua_+\da_-}$. In this example we only consider states with one electron confined to each dot, and therefore omit the dot numbers in our notation for the sake of notational brevity. Now perform the following operations to obtain $\ket{\Psi_{\text{valley}}}$:

\begin{align*}
\ket{\Psi_{\text{valley}}}&=\sqrt{SWAP_{23}}\sqrt{SWAP_{12}}\sqrt{SWAP_{34}}\\&\times\sqrt{SWAP_{23}}\ket{\ua_-\da_+\ua_+\da_-}
\numb
\label{eqn:seq1}
\end{align*} 

Then $\ket{\Psi_{\text{valley}}}$ can be explicitly obtained, and is given by:

\begin{align*}
&\ket{\Psi_{\text{valley}}}=\frac{1}{4}\Big[\ket{\ua_+\ua_-\da_+\da_-}-i\ket{\ua_+\ua_-\da_-\da_+}\\&+2i\ket{\ua_-\ua_+\da_+\da_-}+\ket{\ua_-\ua_+\da_-\da_+}-\ket{\ua_+\da_-\ua_-\da_+}\\&-i\ket{\ua_+\da_+\ua_-\da_-}-i\ket{\ua_-\da_-\ua_+\da_+}+\ket{\da_+\ua_-\da_-\ua_+}\\&-i\ket{\da_+\da_-\ua_-\ua_+}+\ket{\ua_-\da_-\da_+\ua_+}+i\ket{\ua_-\da_+\da_-\ua_+}\\&+i\ket{\da_+\ua_-\ua_+\da_-}+\ket{\da_+\ua_+\ua_-\da_-}\Big]
\numb
\end{align*}

From this state, we calculate the measurement probability of the second dot being measured to be spin up, obtaining $P(2_\ua)=5/8$. Now consider the ideal one-valley case, where the same set of operations are performed on a system of four spins in the same initial spin configuration. Then the resulting state $\ket{\Psi_{\text{ideal}}}$ will be given by:

\begin{align*}
\ket{\Psi_{\text{ideal}}}&=\frac{1}{4}\Big[(2+i)\ket{\ua\ua\da\da}-(1+2i)\ket{\ua\da\ua\da}+\ket{\da\ua\da\ua}\\&-i\ket{\da\da\ua\ua}+(1+i)\ket{\ua\da\da\ua}+(1+i)\ket{\da\ua\ua\da}\Big]
\numb
\end{align*}

In the ideal case, the probability of measuring the second spin to be up is $P(2_\ua)=1/2$, which is different from the case above where two valley states are present. This discrepancy arises from the fact that in the presence of valley states, there is a distinction between certain states that would be considered identical in the ideal case (e.g. $\ket{\ua_+\ua_-\da_+\da_-}$ and $\ket{\ua_-\ua_+\da_-\da_+}$). This distinction prevents constructive or destructive interference between the states, which influences the final measurement probabilities (in the ideal case both states are $\ket{\ua\ua\da\da}$ so their amplitudes should add together). This may have important implications for Si qubits. In order to perform correct calculations, all dots must be initialized to the same valley state, and the opposite valley state should be considered a leakage state. The presence of a phase difference between dots $\phi_2-\phi_1$ (leading to a $t_-$ term) also introduces leakage into the system. However, the leakage states behave almost like the ideal states for small systems, and in fact they are indistinguishable for systems of only two dots, as we have shown in the previous section. This makes errors resulting from leakage difficult to detect in small systems and small gate sequences. Thus fidelities of two-qubit gates measured in two-qubit systems may be measured as higher than their true value in multiqubit circuits, because these measurements cannot account for leakage into other valley states without applying a larger sequence of gates such as that of Eq. (\ref{eqn:seq1}) to a larger number of dots.

\begin{figure}[!htb]
	\centering
	\includegraphics[width=.5\columnwidth]{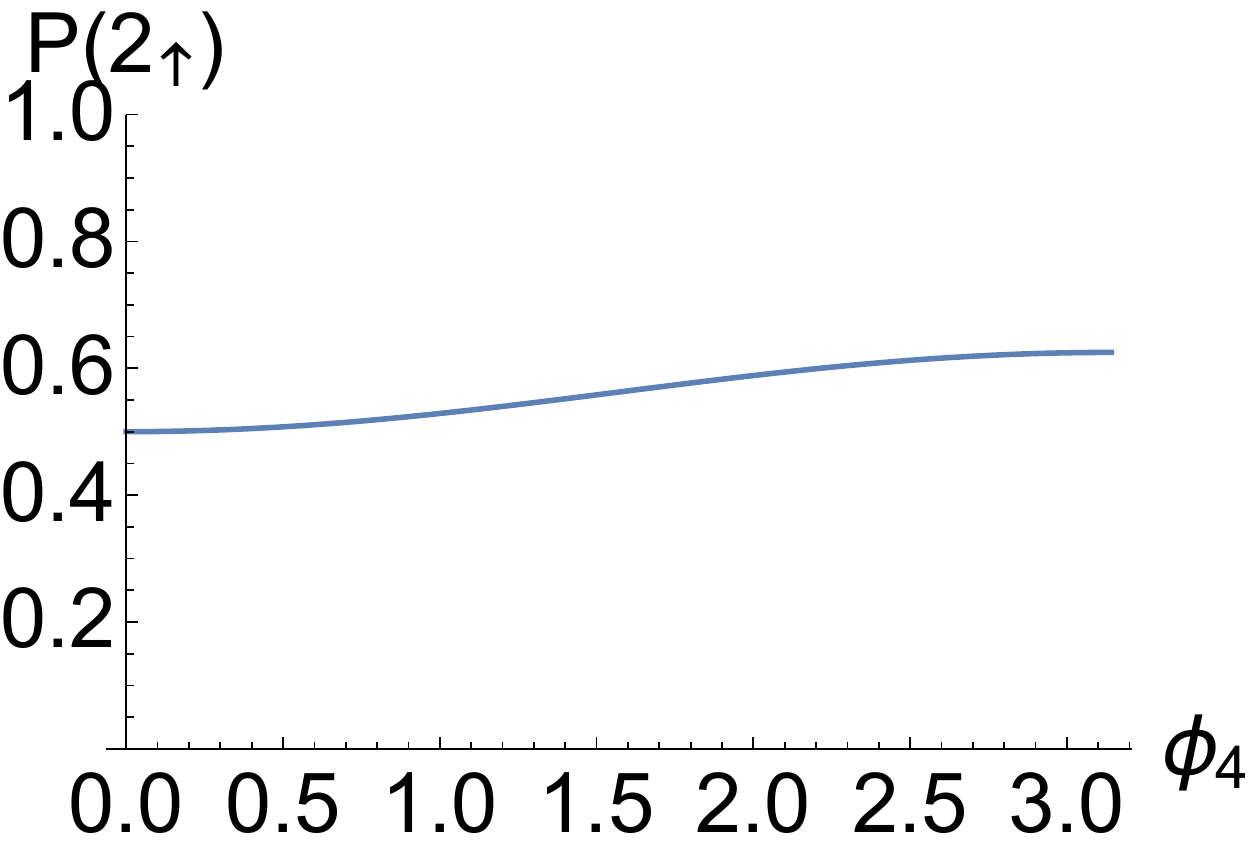}
	\caption{Probability that dot 2 will be measured as $\ua$ after the sequence of gates shown in eq. (\ref{eqn:seq1}) is applied to the initial state $\ket{\ua_+\da_+\ua_+\da_+}$. Here $\Delta_i=0$, $\phi_1=\phi_2=\phi_3=0$, and $\phi_4$ varies.}
	\label{fig:seq1}
\end{figure}

The sequence of gates given in Eq. (\ref{eqn:seq1}) is a short sequence of gates which yields a different measurement outcome in an ideal system than in a system with multiple valley states. This effect is present for any initial state where the electrons with the same spin do not all have the same valley state. Note that the initial state is dependent on the values of $\phi_i$, and the only way to control them is by controlling the valley phase. In Fig. \ref{fig:seq1} we plot the measurement probability of $\ket{2_\ua}$ versus the value of $\phi_4$, with $\phi_1=\phi_2=\phi_3=0$. When $\phi_4$ also equals 0, the initial valley states of all electrons are the same, and the measurement probability is the same as in the ideal case. As $\phi_4$ varies away from 0, the measurement probability increases. This sequence of gates can be used to demonstrate the presence of valley-induced error, as if the state $\ket{2_\ua}$ is measured with probability greater than 1/2, this is a result of valley-induced error. However this sequence of gates cannot be used to show the converse in noisy systems, because if $\ket{2_\ua}$ is measured with probability 1/2, the result is indistinguishable from noise-induced decoherence.

We now give a different sequence of gates which has a measurement probability of 1 in the ideal case, and thus can be used to demonstrate initialization of electrons in the same valley state. Consider a ring of 4 quantum dots with the initial state $\ket{\ua_-\da_+\ua_+\da_-}$. Perform the following sequence of $\sqrt{SWAP}$ gates:

\begin{align*}
&\ket{\Psi_{\text{valley}}}=\sqrt{SWAP_{23}}\sqrt{SWAP_{14}}\sqrt{SWAP_{13}}\\&\times\sqrt{SWAP_{24}}\sqrt{SWAP_{12}}\sqrt{SWAP_{34}}\ket{\ua_-\da_+\ua_+\da_-}
\label{eqn:seq2}
\end{align*} 

For this sequence of gates, $\ket{\Psi_{\text{valley}}}$ is given by:

\begin{align*}
&\ket{\Psi_{\text{valley}}}=\frac{1}{4}\Big[\ket{\ua_+\ua_-\da_-\da_+}-\ket{\ua_-\ua_+\da_+\da_-}\\&+\ket{\da_+\da_-\ua_-\ua_+}-\ket{\da_-\da_+\ua_+\ua_-}+\ket{\ua_+\da_-\da_+\ua_-}\\&-\ket{\ua_-\da_+\da_-\ua_+}-\ket{\da_+\ua_-\ua_+\da_-}+\ket{\da_-\ua_+\ua_-\da_+}\\&+2\ket{\da_+\ua_+\da_-\ua_-}+2\ket{\da_-\ua_-\da_+\ua_+}\Big]
\numb
\end{align*}

The corresponding ideal state $\ket{\Psi_{\text{ideal}}}$ is simply:

\begin{align*}
\ket{\Psi_{\text{ideal}}}=\ket{\da\ua\da\ua}
\numb
\end{align*}

With this sequence of gates, the ideal single-valley case will have measurement probabilities of $100\%$, but the corresponding two-valley case will only have corresponding measurement probabilities of $75\%$, simulating an unknown apparent decoherence although it arises simply from the inevitable spin-valley entanglement which is omnipresent in the multi-valley qubits. 

\begin{figure}[!htb]
	\centering
	\includegraphics[width=.5\columnwidth]{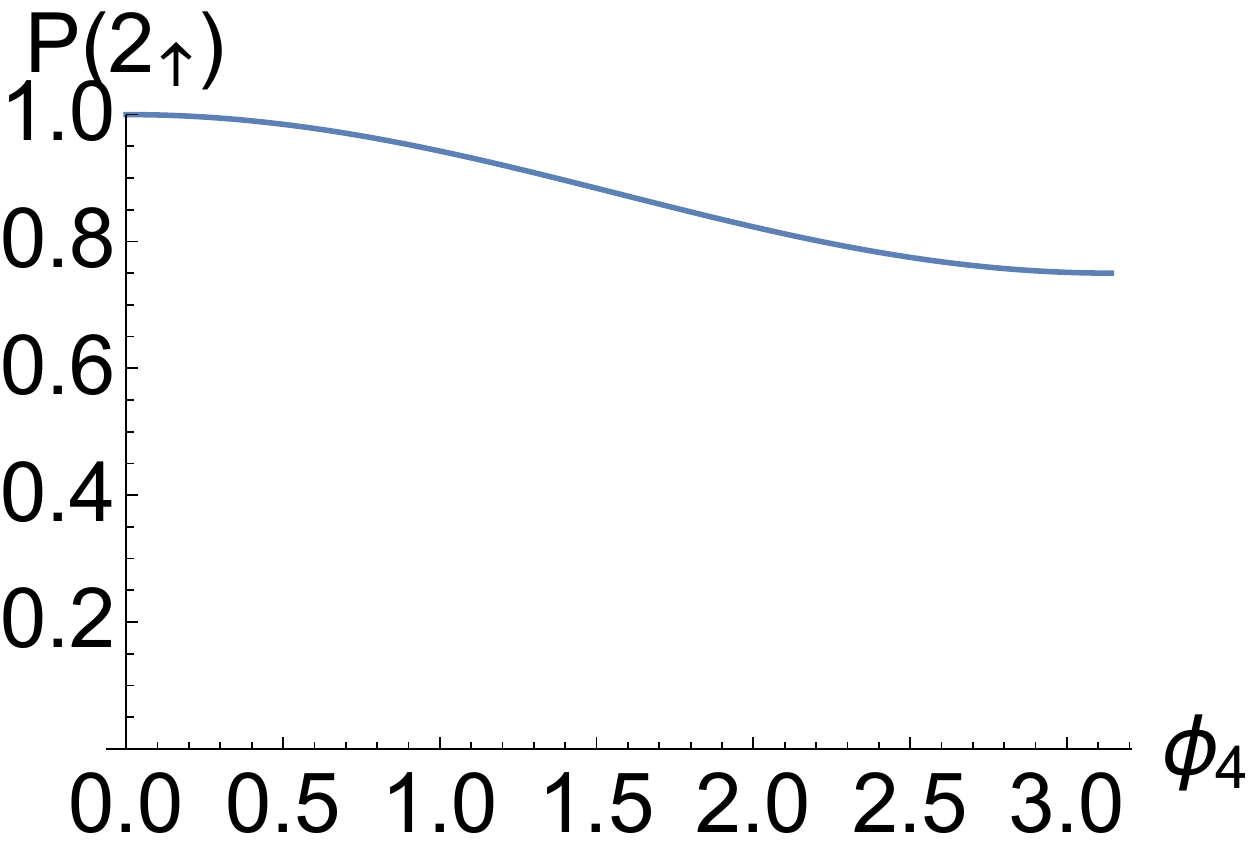}
	\caption{Probability that dot 2 will be measured as $\ua$ after the sequence of gates shown in eq. (\ref{eqn:seq2}) is applied to the initial state $\ket{\ua_+\da_+\ua_+\da_+}$. Here $\Delta_i=0$, $\phi_1=\phi_2=\phi_3=0$, and $\phi_4$ varies.}
\end{figure}

\begin{figure}[!htb]
	\centering
	\includegraphics[width=.5\columnwidth]{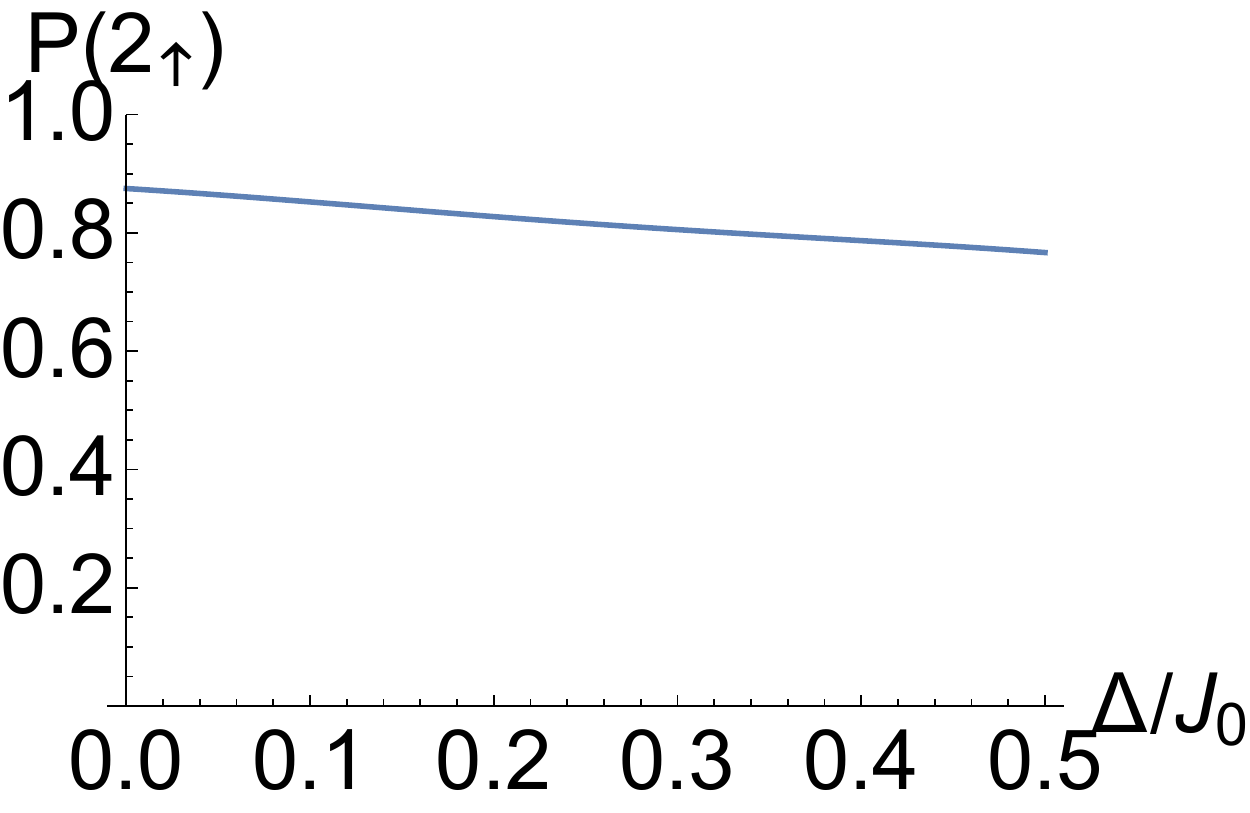}
	\caption{Probability that dot 2 will be measured as $\ua$ after the sequence of gates shown in eq. (\ref{eqn:seq2}) is applied to the initial state $\ket{\ua_+\da_+\ua_+\da_+}$. Here $\phi_1=\phi_2=\phi_3=0$, $\phi_4=\pi/2$, and $\Delta$ varies.}
\end{figure}

\section{Conclusion}

We used an effective Hubbard model to investigate the effects of valley states in Silicon quantum dots in the context of exchange gate operations in multiqubit systems at zero temperature. We first considered a system of two exchange-coupled quantum dots, and determined the eigenstates and energies for both the singlet and triplet spin configurations to leading order in $t_c^2/U$. We considered the limits where $\Delta\gg J_0$ and $\Delta\ll J_0$, and found that in both cases the singlet and triplet spin configurations share the same eigenstates, though their energies differ. When the valley splitting is large, any state not initialized to valley eigenstates will lead to spin-valley entanglement as the system evolves. However, as long as the valley splitting is large and the system is initialized to the valley ground state, the evolution of the system will not be affected to leading order in $\Delta/J_0$. Thus valley degrees of freedom are not problematic for exchange gate operations as long as the valley splitting $\Delta$ is sufficiently large. For small valley splitting or completely degenerate valley states, spin-valley entanglement will arise any time the electrons are not initialized to the same valley state. Without the ability to perform valley state measurements, spin-valley entanglement will obfuscate any information stored in the electron spin states, making it impossible to perform quantum computations in such a state. For a two qubit system with $\Delta\ll J_0$, if all spins are initialized in $z$-eigenstates and measured in the $z$-basis, then the resulting measurement probabilities will be the same as an ideal single-valley system. However, this does not extend to systems with more qubits, as spin-valley entanglement can introduce errors despite states being initialized and measured in the $z$-basis. This means that two-qubit gate fidelities measured by performing a single gate in a two-qubit system will give deceptively high fidelity results, because the presence of valley states does not affect the fidelity measurement even though it can affect operations in systems with more qubits. To demonstrate robustness to spin-valley entanglement without the ability to measure valley states themselves requires a longer series of gates in a system with more qubits. We emphasize that our use of the Hubbard model is not an approximation here since the coupled spin qubit system is indeed equivalent to the Hubbard model where the Hubbard interaction $U$ is simply related to the exchange coupling $J_0$ between the dots through $U=t_c^2/J_0$, where $t_c$ is the inter-dot hopping energy from wavefunction overlap.  Our fundamental finding of an apparent quantum leakage (or effective decoherence) due to spin-valley entanglement is also independent of additional complications arising from varying exchange couplings and/or inter-dot hopping through the circuit-- all they do is to complicate the expressions for the leakage, but the basic physics of spin-valley entanglement remains the same.

We note that our considerations on spin-valley entanglement apply equally well to all-exchange gate operations considered in Ref. \onlinecite{DiVincenzoNAT2000}-- anytime the inter-dot exchange coupling is used to carry out gate operations, the valley-spin entanglement (and potential decoherence to valley states) studied in this work becomes relevant.  Right now, the Si qubit platforms have rather small exchange coupling, and the problem discussed in this work is most likely not crucial to the current generation of few qubit systems with rather small exchange coupling values.  But faster gate operations in multiqubit circuits will necessitate larger exchange coupling strength in the future, making our dynamical consideration relevant as one must ensure that not only is the valley splitting much larger than the electron temperature in the qubits, it is also much larger than the inter-dot exchange coupling used in the gate operations.

\acknowledgements

This work is supported by the Laboratory for Physical Sciences.

\bibliography{valleybib}

\begin{thebibliography}{28}
\expandafter\ifx\csname natexlab\endcsname\relax\def\natexlab#1{#1}\fi
\expandafter\ifx\csname bibnamefont\endcsname\relax
  \def\bibnamefont#1{#1}\fi
\expandafter\ifx\csname bibfnamefont\endcsname\relax
  \def\bibfnamefont#1{#1}\fi
\expandafter\ifx\csname citenamefont\endcsname\relax
  \def\citenamefont#1{#1}\fi
\expandafter\ifx\csname url\endcsname\relax
  \def\url#1{\texttt{#1}}\fi
\expandafter\ifx\csname urlprefix\endcsname\relax\def\urlprefix{URL }\fi
\providecommand{\bibinfo}[2]{#2}
\providecommand{\eprint}[2][]{\url{#2}}

\bibitem[{\citenamefont{Huang et~al.}(2019)\citenamefont{Huang, Yang, Chan,
  Tanttu, Hensen, Leon, Fogarty, Hwang, Hudson, Itoh et~al.}}]{HuangNat2019}
\bibinfo{author}{\bibfnamefont{W.}~\bibnamefont{Huang}},
  \bibinfo{author}{\bibfnamefont{C.~H.} \bibnamefont{Yang}},
  \bibinfo{author}{\bibfnamefont{K.~W.} \bibnamefont{Chan}},
  \bibinfo{author}{\bibfnamefont{T.}~\bibnamefont{Tanttu}},
  \bibinfo{author}{\bibfnamefont{B.}~\bibnamefont{Hensen}},
  \bibinfo{author}{\bibfnamefont{R.~C.~C.} \bibnamefont{Leon}},
  \bibinfo{author}{\bibfnamefont{M.~A.} \bibnamefont{Fogarty}},
  \bibinfo{author}{\bibfnamefont{J.~C.~C.} \bibnamefont{Hwang}},
  \bibinfo{author}{\bibfnamefont{F.~E.} \bibnamefont{Hudson}},
  \bibinfo{author}{\bibfnamefont{K.~M.} \bibnamefont{Itoh}},
  \bibnamefont{et~al.}, \bibinfo{journal}{Nature}
  \textbf{\bibinfo{volume}{569}}, \bibinfo{pages}{532} (\bibinfo{year}{2019}),
  ISSN \bibinfo{issn}{1476-4687},
  \urlprefix\url{https://doi.org/10.1038/s41586-019-1197-0}.

\bibitem[{\citenamefont{Maurand et~al.}(2016)\citenamefont{Maurand, Jehl,
  Kotekar-Patil, Corna, Bohuslavskyi, Lavi{\'e}ville, Hutin, Barraud, Vinet,
  Sanquer et~al.}}]{MaurandNC2016}
\bibinfo{author}{\bibfnamefont{R.}~\bibnamefont{Maurand}},
  \bibinfo{author}{\bibfnamefont{X.}~\bibnamefont{Jehl}},
  \bibinfo{author}{\bibfnamefont{D.}~\bibnamefont{Kotekar-Patil}},
  \bibinfo{author}{\bibfnamefont{A.}~\bibnamefont{Corna}},
  \bibinfo{author}{\bibfnamefont{H.}~\bibnamefont{Bohuslavskyi}},
  \bibinfo{author}{\bibfnamefont{R.}~\bibnamefont{Lavi{\'e}ville}},
  \bibinfo{author}{\bibfnamefont{L.}~\bibnamefont{Hutin}},
  \bibinfo{author}{\bibfnamefont{S.}~\bibnamefont{Barraud}},
  \bibinfo{author}{\bibfnamefont{M.}~\bibnamefont{Vinet}},
  \bibinfo{author}{\bibfnamefont{M.}~\bibnamefont{Sanquer}},
  \bibnamefont{et~al.}, \bibinfo{journal}{Nature Communications}
  \textbf{\bibinfo{volume}{7}}, \bibinfo{pages}{13575} (\bibinfo{year}{2016}),
  ISSN \bibinfo{issn}{2041-1723},
  \urlprefix\url{https://doi.org/10.1038/ncomms13575}.

\bibitem[{\citenamefont{Zhao et~al.}(2019)\citenamefont{Zhao, Tanttu, Tan,
  Hensen, Chan, Hwang, Leon, Yang, Gilbert, Hudson et~al.}}]{ZhaoNC2019}
\bibinfo{author}{\bibfnamefont{R.}~\bibnamefont{Zhao}},
  \bibinfo{author}{\bibfnamefont{T.}~\bibnamefont{Tanttu}},
  \bibinfo{author}{\bibfnamefont{K.~Y.} \bibnamefont{Tan}},
  \bibinfo{author}{\bibfnamefont{B.}~\bibnamefont{Hensen}},
  \bibinfo{author}{\bibfnamefont{K.~W.} \bibnamefont{Chan}},
  \bibinfo{author}{\bibfnamefont{J.~C.~C.} \bibnamefont{Hwang}},
  \bibinfo{author}{\bibfnamefont{R.~C.~C.} \bibnamefont{Leon}},
  \bibinfo{author}{\bibfnamefont{C.~H.} \bibnamefont{Yang}},
  \bibinfo{author}{\bibfnamefont{W.}~\bibnamefont{Gilbert}},
  \bibinfo{author}{\bibfnamefont{F.~E.} \bibnamefont{Hudson}},
  \bibnamefont{et~al.}, \bibinfo{journal}{Nature Communications}
  \textbf{\bibinfo{volume}{10}}, \bibinfo{pages}{5500} (\bibinfo{year}{2019}),
  ISSN \bibinfo{issn}{2041-1723},
  \urlprefix\url{https://doi.org/10.1038/s41467-019-13416-7}.

\bibitem[{\citenamefont{Yang et~al.}(2020)\citenamefont{Yang, Leon, Hwang,
  Saraiva, Tanttu, Huang, Camirand~Lemyre, Chan, Tan, Hudson
  et~al.}}]{YangNat2020}
\bibinfo{author}{\bibfnamefont{C.~H.} \bibnamefont{Yang}},
  \bibinfo{author}{\bibfnamefont{R.~C.~C.} \bibnamefont{Leon}},
  \bibinfo{author}{\bibfnamefont{J.~C.~C.} \bibnamefont{Hwang}},
  \bibinfo{author}{\bibfnamefont{A.}~\bibnamefont{Saraiva}},
  \bibinfo{author}{\bibfnamefont{T.}~\bibnamefont{Tanttu}},
  \bibinfo{author}{\bibfnamefont{W.}~\bibnamefont{Huang}},
  \bibinfo{author}{\bibfnamefont{J.}~\bibnamefont{Camirand~Lemyre}},
  \bibinfo{author}{\bibfnamefont{K.~W.} \bibnamefont{Chan}},
  \bibinfo{author}{\bibfnamefont{K.~Y.} \bibnamefont{Tan}},
  \bibinfo{author}{\bibfnamefont{F.~E.} \bibnamefont{Hudson}},
  \bibnamefont{et~al.}, \bibinfo{journal}{Nature}
  \textbf{\bibinfo{volume}{580}}, \bibinfo{pages}{350} (\bibinfo{year}{2020}),
  ISSN \bibinfo{issn}{1476-4687},
  \urlprefix\url{https://doi.org/10.1038/s41586-020-2171-6}.

\bibitem[{\citenamefont{Petit et~al.}(2020)\citenamefont{Petit, Eenink, Russ,
  Lawrie, Hendrickx, Philips, Clarke, Vandersypen, and
  Veldhorst}}]{PetitNat2020}
\bibinfo{author}{\bibfnamefont{L.}~\bibnamefont{Petit}},
  \bibinfo{author}{\bibfnamefont{H.~G.~J.} \bibnamefont{Eenink}},
  \bibinfo{author}{\bibfnamefont{M.}~\bibnamefont{Russ}},
  \bibinfo{author}{\bibfnamefont{W.~I.~L.} \bibnamefont{Lawrie}},
  \bibinfo{author}{\bibfnamefont{N.~W.} \bibnamefont{Hendrickx}},
  \bibinfo{author}{\bibfnamefont{S.~G.~J.} \bibnamefont{Philips}},
  \bibinfo{author}{\bibfnamefont{J.~S.} \bibnamefont{Clarke}},
  \bibinfo{author}{\bibfnamefont{L.~M.~K.} \bibnamefont{Vandersypen}},
  \bibnamefont{and}
  \bibinfo{author}{\bibfnamefont{M.}~\bibnamefont{Veldhorst}},
  \bibinfo{journal}{Nature} \textbf{\bibinfo{volume}{580}},
  \bibinfo{pages}{355} (\bibinfo{year}{2020}), ISSN \bibinfo{issn}{1476-4687},
  \urlprefix\url{https://doi.org/10.1038/s41586-020-2170-7}.

\bibitem[{\citenamefont{Fogarty et~al.}(2018)\citenamefont{Fogarty, Chan,
  Hensen, Huang, Tanttu, Yang, Laucht, Veldhorst, Hudson, Itoh
  et~al.}}]{FogartyNC2018}
\bibinfo{author}{\bibfnamefont{M.~A.} \bibnamefont{Fogarty}},
  \bibinfo{author}{\bibfnamefont{K.~W.} \bibnamefont{Chan}},
  \bibinfo{author}{\bibfnamefont{B.}~\bibnamefont{Hensen}},
  \bibinfo{author}{\bibfnamefont{W.}~\bibnamefont{Huang}},
  \bibinfo{author}{\bibfnamefont{T.}~\bibnamefont{Tanttu}},
  \bibinfo{author}{\bibfnamefont{C.~H.} \bibnamefont{Yang}},
  \bibinfo{author}{\bibfnamefont{A.}~\bibnamefont{Laucht}},
  \bibinfo{author}{\bibfnamefont{M.}~\bibnamefont{Veldhorst}},
  \bibinfo{author}{\bibfnamefont{F.~E.} \bibnamefont{Hudson}},
  \bibinfo{author}{\bibfnamefont{K.~M.} \bibnamefont{Itoh}},
  \bibnamefont{et~al.}, \bibinfo{journal}{Nature Communications}
  \textbf{\bibinfo{volume}{9}}, \bibinfo{pages}{4370} (\bibinfo{year}{2018}),
  ISSN \bibinfo{issn}{2041-1723},
  \urlprefix\url{https://doi.org/10.1038/s41467-018-06039-x}.

\bibitem[{\citenamefont{Veldhorst et~al.}(2015)\citenamefont{Veldhorst, Yang,
  Hwang, Huang, Dehollain, Muhonen, Simmons, Laucht, Hudson, Itoh
  et~al.}}]{VeldhorstNat2015}
\bibinfo{author}{\bibfnamefont{M.}~\bibnamefont{Veldhorst}},
  \bibinfo{author}{\bibfnamefont{C.~H.} \bibnamefont{Yang}},
  \bibinfo{author}{\bibfnamefont{J.~C.~C.} \bibnamefont{Hwang}},
  \bibinfo{author}{\bibfnamefont{W.}~\bibnamefont{Huang}},
  \bibinfo{author}{\bibfnamefont{J.~P.} \bibnamefont{Dehollain}},
  \bibinfo{author}{\bibfnamefont{J.~T.} \bibnamefont{Muhonen}},
  \bibinfo{author}{\bibfnamefont{S.}~\bibnamefont{Simmons}},
  \bibinfo{author}{\bibfnamefont{A.}~\bibnamefont{Laucht}},
  \bibinfo{author}{\bibfnamefont{F.~E.} \bibnamefont{Hudson}},
  \bibinfo{author}{\bibfnamefont{K.~M.} \bibnamefont{Itoh}},
  \bibnamefont{et~al.}, \bibinfo{journal}{Nature}
  \textbf{\bibinfo{volume}{526}}, \bibinfo{pages}{410} (\bibinfo{year}{2015}),
  ISSN \bibinfo{issn}{1476-4687},
  \urlprefix\url{https://doi.org/10.1038/nature15263}.

\bibitem[{\citenamefont{Zajac et~al.}(2018)\citenamefont{Zajac, Sigillito,
  Russ, Borjans, Taylor, Burkard, and Petta}}]{ZajacSci2018}
\bibinfo{author}{\bibfnamefont{D.~M.} \bibnamefont{Zajac}},
  \bibinfo{author}{\bibfnamefont{A.~J.} \bibnamefont{Sigillito}},
  \bibinfo{author}{\bibfnamefont{M.}~\bibnamefont{Russ}},
  \bibinfo{author}{\bibfnamefont{F.}~\bibnamefont{Borjans}},
  \bibinfo{author}{\bibfnamefont{J.~M.} \bibnamefont{Taylor}},
  \bibinfo{author}{\bibfnamefont{G.}~\bibnamefont{Burkard}}, \bibnamefont{and}
  \bibinfo{author}{\bibfnamefont{J.~R.} \bibnamefont{Petta}},
  \bibinfo{journal}{Science} \textbf{\bibinfo{volume}{359}},
  \bibinfo{pages}{439} (\bibinfo{year}{2018}), ISSN \bibinfo{issn}{0036-8075},
  \eprint{https://science.sciencemag.org/content/359/6374/439.full.pdf},
  \urlprefix\url{https://science.sciencemag.org/content/359/6374/439}.

\bibitem[{\citenamefont{Watson et~al.}(2018)\citenamefont{Watson, Philips,
  Kawakami, Ward, Scarlino, Veldhorst, Savage, Lagally, Friesen, Coppersmith
  et~al.}}]{WatsonNat2018}
\bibinfo{author}{\bibfnamefont{T.~F.} \bibnamefont{Watson}},
  \bibinfo{author}{\bibfnamefont{S.~G.~J.} \bibnamefont{Philips}},
  \bibinfo{author}{\bibfnamefont{E.}~\bibnamefont{Kawakami}},
  \bibinfo{author}{\bibfnamefont{D.~R.} \bibnamefont{Ward}},
  \bibinfo{author}{\bibfnamefont{P.}~\bibnamefont{Scarlino}},
  \bibinfo{author}{\bibfnamefont{M.}~\bibnamefont{Veldhorst}},
  \bibinfo{author}{\bibfnamefont{D.~E.} \bibnamefont{Savage}},
  \bibinfo{author}{\bibfnamefont{M.~G.} \bibnamefont{Lagally}},
  \bibinfo{author}{\bibfnamefont{M.}~\bibnamefont{Friesen}},
  \bibinfo{author}{\bibfnamefont{S.~N.} \bibnamefont{Coppersmith}},
  \bibnamefont{et~al.}, \bibinfo{journal}{Nature}
  \textbf{\bibinfo{volume}{555}}, \bibinfo{pages}{633} (\bibinfo{year}{2018}),
  ISSN \bibinfo{issn}{1476-4687},
  \urlprefix\url{https://doi.org/10.1038/nature25766}.

\bibitem[{\citenamefont{Zajac et~al.}(2016)\citenamefont{Zajac, Hazard, Mi,
  Nielsen, and Petta}}]{ZajacPRApp2016}
\bibinfo{author}{\bibfnamefont{D.~M.} \bibnamefont{Zajac}},
  \bibinfo{author}{\bibfnamefont{T.~M.} \bibnamefont{Hazard}},
  \bibinfo{author}{\bibfnamefont{X.}~\bibnamefont{Mi}},
  \bibinfo{author}{\bibfnamefont{E.}~\bibnamefont{Nielsen}}, \bibnamefont{and}
  \bibinfo{author}{\bibfnamefont{J.~R.} \bibnamefont{Petta}},
  \bibinfo{journal}{Phys. Rev. Applied} \textbf{\bibinfo{volume}{6}},
  \bibinfo{pages}{054013} (\bibinfo{year}{2016}),
  \urlprefix\url{https://link.aps.org/doi/10.1103/PhysRevApplied.6.054013}.

\bibitem[{\citenamefont{Mills et~al.}(2019)\citenamefont{Mills, Zajac, Gullans,
  Schupp, Hazard, and Petta}}]{MillsNC2019}
\bibinfo{author}{\bibfnamefont{A.~R.} \bibnamefont{Mills}},
  \bibinfo{author}{\bibfnamefont{D.~M.} \bibnamefont{Zajac}},
  \bibinfo{author}{\bibfnamefont{M.~J.} \bibnamefont{Gullans}},
  \bibinfo{author}{\bibfnamefont{F.~J.} \bibnamefont{Schupp}},
  \bibinfo{author}{\bibfnamefont{T.~M.} \bibnamefont{Hazard}},
  \bibnamefont{and} \bibinfo{author}{\bibfnamefont{J.~R.} \bibnamefont{Petta}},
  \bibinfo{journal}{Nature Communications} \textbf{\bibinfo{volume}{10}},
  \bibinfo{pages}{1063} (\bibinfo{year}{2019}), ISSN \bibinfo{issn}{2041-1723},
  \urlprefix\url{https://doi.org/10.1038/s41467-019-08970-z}.

\bibitem[{\citenamefont{Borjans et~al.}(2020)\citenamefont{Borjans, Croot, Mi,
  Gullans, and Petta}}]{BorjansNat2020}
\bibinfo{author}{\bibfnamefont{F.}~\bibnamefont{Borjans}},
  \bibinfo{author}{\bibfnamefont{X.~G.} \bibnamefont{Croot}},
  \bibinfo{author}{\bibfnamefont{X.}~\bibnamefont{Mi}},
  \bibinfo{author}{\bibfnamefont{M.~J.} \bibnamefont{Gullans}},
  \bibnamefont{and} \bibinfo{author}{\bibfnamefont{J.~R.} \bibnamefont{Petta}},
  \bibinfo{journal}{Nature} \textbf{\bibinfo{volume}{577}},
  \bibinfo{pages}{195} (\bibinfo{year}{2020}), ISSN \bibinfo{issn}{1476-4687},
  \urlprefix\url{https://doi.org/10.1038/s41586-019-1867-y}.

\bibitem[{\citenamefont{Sigillito et~al.}(2019)\citenamefont{Sigillito,
  Gullans, Edge, Borselli, and Petta}}]{SigillitoNPJQI2019}
\bibinfo{author}{\bibfnamefont{A.~J.} \bibnamefont{Sigillito}},
  \bibinfo{author}{\bibfnamefont{M.~J.} \bibnamefont{Gullans}},
  \bibinfo{author}{\bibfnamefont{L.~F.} \bibnamefont{Edge}},
  \bibinfo{author}{\bibfnamefont{M.}~\bibnamefont{Borselli}}, \bibnamefont{and}
  \bibinfo{author}{\bibfnamefont{J.~R.} \bibnamefont{Petta}},
  \bibinfo{journal}{npj Quantum Information} \textbf{\bibinfo{volume}{5}},
  \bibinfo{pages}{110} (\bibinfo{year}{2019}), ISSN \bibinfo{issn}{2056-6387},
  \urlprefix\url{https://doi.org/10.1038/s41534-019-0225-0}.

\bibitem[{\citenamefont{Xue et~al.}(2020)\citenamefont{Xue, D'Anjou, Watson,
  Ward, Savage, Lagally, Friesen, Coppersmith, Eriksson, Coish
  et~al.}}]{XuePRX2020}
\bibinfo{author}{\bibfnamefont{X.}~\bibnamefont{Xue}},
  \bibinfo{author}{\bibfnamefont{B.}~\bibnamefont{D'Anjou}},
  \bibinfo{author}{\bibfnamefont{T.~F.} \bibnamefont{Watson}},
  \bibinfo{author}{\bibfnamefont{D.~R.} \bibnamefont{Ward}},
  \bibinfo{author}{\bibfnamefont{D.~E.} \bibnamefont{Savage}},
  \bibinfo{author}{\bibfnamefont{M.~G.} \bibnamefont{Lagally}},
  \bibinfo{author}{\bibfnamefont{M.}~\bibnamefont{Friesen}},
  \bibinfo{author}{\bibfnamefont{S.~N.} \bibnamefont{Coppersmith}},
  \bibinfo{author}{\bibfnamefont{M.~A.} \bibnamefont{Eriksson}},
  \bibinfo{author}{\bibfnamefont{W.~A.} \bibnamefont{Coish}},
  \bibnamefont{et~al.}, \bibinfo{journal}{Phys. Rev. X}
  \textbf{\bibinfo{volume}{10}}, \bibinfo{pages}{021006}
  (\bibinfo{year}{2020}),
  \urlprefix\url{https://link.aps.org/doi/10.1103/PhysRevX.10.021006}.

\bibitem[{\citenamefont{Kawakami et~al.}(2014)\citenamefont{Kawakami, Scarlino,
  Ward, Braakman, Savage, Lagally, Friesen, Coppersmith, Eriksson, and
  Vandersypen}}]{KawakamiNN2014}
\bibinfo{author}{\bibfnamefont{E.}~\bibnamefont{Kawakami}},
  \bibinfo{author}{\bibfnamefont{P.}~\bibnamefont{Scarlino}},
  \bibinfo{author}{\bibfnamefont{D.~R.} \bibnamefont{Ward}},
  \bibinfo{author}{\bibfnamefont{F.~R.} \bibnamefont{Braakman}},
  \bibinfo{author}{\bibfnamefont{D.~E.} \bibnamefont{Savage}},
  \bibinfo{author}{\bibfnamefont{M.~G.} \bibnamefont{Lagally}},
  \bibinfo{author}{\bibfnamefont{M.}~\bibnamefont{Friesen}},
  \bibinfo{author}{\bibfnamefont{S.~N.} \bibnamefont{Coppersmith}},
  \bibinfo{author}{\bibfnamefont{M.~A.} \bibnamefont{Eriksson}},
  \bibnamefont{and} \bibinfo{author}{\bibfnamefont{L.~M.~K.}
  \bibnamefont{Vandersypen}}, \bibinfo{journal}{Nature Nanotechnology}
  \textbf{\bibinfo{volume}{9}}, \bibinfo{pages}{666} (\bibinfo{year}{2014}),
  ISSN \bibinfo{issn}{1748-3395},
  \urlprefix\url{https://doi.org/10.1038/nnano.2014.153}.

\bibitem[{\citenamefont{Xue et~al.}(2019)\citenamefont{Xue, Watson, Helsen,
  Ward, Savage, Lagally, Coppersmith, Eriksson, Wehner, and
  Vandersypen}}]{XuePRX2019}
\bibinfo{author}{\bibfnamefont{X.}~\bibnamefont{Xue}},
  \bibinfo{author}{\bibfnamefont{T.~F.} \bibnamefont{Watson}},
  \bibinfo{author}{\bibfnamefont{J.}~\bibnamefont{Helsen}},
  \bibinfo{author}{\bibfnamefont{D.~R.} \bibnamefont{Ward}},
  \bibinfo{author}{\bibfnamefont{D.~E.} \bibnamefont{Savage}},
  \bibinfo{author}{\bibfnamefont{M.~G.} \bibnamefont{Lagally}},
  \bibinfo{author}{\bibfnamefont{S.~N.} \bibnamefont{Coppersmith}},
  \bibinfo{author}{\bibfnamefont{M.~A.} \bibnamefont{Eriksson}},
  \bibinfo{author}{\bibfnamefont{S.}~\bibnamefont{Wehner}}, \bibnamefont{and}
  \bibinfo{author}{\bibfnamefont{L.~M.~K.} \bibnamefont{Vandersypen}},
  \bibinfo{journal}{Phys. Rev. X} \textbf{\bibinfo{volume}{9}},
  \bibinfo{pages}{021011} (\bibinfo{year}{2019}),
  \urlprefix\url{https://link.aps.org/doi/10.1103/PhysRevX.9.021011}.

\bibitem[{\citenamefont{Shor}(1997)}]{ShorSIAMJC1997}
\bibinfo{author}{\bibfnamefont{P.~W.} \bibnamefont{Shor}},
  \bibinfo{journal}{SIAM Journal on Computing} \textbf{\bibinfo{volume}{26}},
  \bibinfo{pages}{1484–1509} (\bibinfo{year}{1997}), ISSN
  \bibinfo{issn}{1095-7111},
  \urlprefix\url{http://dx.doi.org/10.1137/S0097539795293172}.

\bibitem[{\citenamefont{Witzel et~al.}(2010)\citenamefont{Witzel, Carroll,
  Morello, Cywi\ifmmode~\acute{n}\else \'{n}\fi{}ski, and
  Das~Sarma}}]{WitzelPRL2010}
\bibinfo{author}{\bibfnamefont{W.~M.} \bibnamefont{Witzel}},
  \bibinfo{author}{\bibfnamefont{M.~S.} \bibnamefont{Carroll}},
  \bibinfo{author}{\bibfnamefont{A.}~\bibnamefont{Morello}},
  \bibinfo{author}{\bibfnamefont{L.}~\bibnamefont{Cywi\ifmmode~\acute{n}\else
  \'{n}\fi{}ski}}, \bibnamefont{and}
  \bibinfo{author}{\bibfnamefont{S.}~\bibnamefont{Das~Sarma}},
  \bibinfo{journal}{Phys. Rev. Lett.} \textbf{\bibinfo{volume}{105}},
  \bibinfo{pages}{187602} (\bibinfo{year}{2010}),
  \urlprefix\url{https://link.aps.org/doi/10.1103/PhysRevLett.105.187602}.

\bibitem[{\citenamefont{Saraiva et~al.}(2009)\citenamefont{Saraiva, Calder\'on,
  Hu, Das~Sarma, and Koiller}}]{SaraivaPRB2009}
\bibinfo{author}{\bibfnamefont{A.~L.} \bibnamefont{Saraiva}},
  \bibinfo{author}{\bibfnamefont{M.~J.} \bibnamefont{Calder\'on}},
  \bibinfo{author}{\bibfnamefont{X.}~\bibnamefont{Hu}},
  \bibinfo{author}{\bibfnamefont{S.}~\bibnamefont{Das~Sarma}},
  \bibnamefont{and} \bibinfo{author}{\bibfnamefont{B.}~\bibnamefont{Koiller}},
  \bibinfo{journal}{Phys. Rev. B} \textbf{\bibinfo{volume}{80}},
  \bibinfo{pages}{081305} (\bibinfo{year}{2009}),
  \urlprefix\url{https://link.aps.org/doi/10.1103/PhysRevB.80.081305}.

\bibitem[{\citenamefont{Saraiva et~al.}(2011)\citenamefont{Saraiva, Calder\'on,
  Capaz, Hu, Das~Sarma, and Koiller}}]{SaraivaPRB2011}
\bibinfo{author}{\bibfnamefont{A.~L.} \bibnamefont{Saraiva}},
  \bibinfo{author}{\bibfnamefont{M.~J.} \bibnamefont{Calder\'on}},
  \bibinfo{author}{\bibfnamefont{R.~B.} \bibnamefont{Capaz}},
  \bibinfo{author}{\bibfnamefont{X.}~\bibnamefont{Hu}},
  \bibinfo{author}{\bibfnamefont{S.}~\bibnamefont{Das~Sarma}},
  \bibnamefont{and} \bibinfo{author}{\bibfnamefont{B.}~\bibnamefont{Koiller}},
  \bibinfo{journal}{Phys. Rev. B} \textbf{\bibinfo{volume}{84}},
  \bibinfo{pages}{155320} (\bibinfo{year}{2011}),
  \urlprefix\url{https://link.aps.org/doi/10.1103/PhysRevB.84.155320}.

\bibitem[{\citenamefont{Culcer et~al.}(2012)\citenamefont{Culcer, Saraiva,
  Koiller, Hu, and Das~Sarma}}]{CulcerPRL2012}
\bibinfo{author}{\bibfnamefont{D.}~\bibnamefont{Culcer}},
  \bibinfo{author}{\bibfnamefont{A.~L.} \bibnamefont{Saraiva}},
  \bibinfo{author}{\bibfnamefont{B.}~\bibnamefont{Koiller}},
  \bibinfo{author}{\bibfnamefont{X.}~\bibnamefont{Hu}}, \bibnamefont{and}
  \bibinfo{author}{\bibfnamefont{S.}~\bibnamefont{Das~Sarma}},
  \bibinfo{journal}{Phys. Rev. Lett.} \textbf{\bibinfo{volume}{108}},
  \bibinfo{pages}{126804} (\bibinfo{year}{2012}),
  \urlprefix\url{https://link.aps.org/doi/10.1103/PhysRevLett.108.126804}.

\bibitem[{\citenamefont{Dodson et~al.}(2021)\citenamefont{Dodson, Ercan,
  Corrigan, Losert, Holman, McJunkin, Edge, Friesen, Coppersmith, and
  Eriksson}}]{DodsonARXIV2021}
\bibinfo{author}{\bibfnamefont{J.~P.} \bibnamefont{Dodson}},
  \bibinfo{author}{\bibfnamefont{H.~E.} \bibnamefont{Ercan}},
  \bibinfo{author}{\bibfnamefont{J.}~\bibnamefont{Corrigan}},
  \bibinfo{author}{\bibfnamefont{M.}~\bibnamefont{Losert}},
  \bibinfo{author}{\bibfnamefont{N.}~\bibnamefont{Holman}},
  \bibinfo{author}{\bibfnamefont{T.}~\bibnamefont{McJunkin}},
  \bibinfo{author}{\bibfnamefont{L.~F.} \bibnamefont{Edge}},
  \bibinfo{author}{\bibfnamefont{M.}~\bibnamefont{Friesen}},
  \bibinfo{author}{\bibfnamefont{S.~N.} \bibnamefont{Coppersmith}},
  \bibnamefont{and} \bibinfo{author}{\bibfnamefont{M.~A.}
  \bibnamefont{Eriksson}} (\bibinfo{year}{2021}),
  \bibinfo{note}{arXiv:2103.14702},
  \urlprefix\url{https://arxiv.org/abs/2103.14702}.

\bibitem[{\citenamefont{Corrigan et~al.}(2020)\citenamefont{Corrigan, Dodson,
  Ercan, Abadillo-Uriel, Thorgrimsson, Knapp, Holman, McJunkin, Neyens,
  MacQuarrie et~al.}}]{CorriganARXIV2020}
\bibinfo{author}{\bibfnamefont{J.}~\bibnamefont{Corrigan}},
  \bibinfo{author}{\bibfnamefont{J.~P.} \bibnamefont{Dodson}},
  \bibinfo{author}{\bibfnamefont{H.~E.} \bibnamefont{Ercan}},
  \bibinfo{author}{\bibfnamefont{J.~C.} \bibnamefont{Abadillo-Uriel}},
  \bibinfo{author}{\bibfnamefont{B.}~\bibnamefont{Thorgrimsson}},
  \bibinfo{author}{\bibfnamefont{T.~J.} \bibnamefont{Knapp}},
  \bibinfo{author}{\bibfnamefont{N.}~\bibnamefont{Holman}},
  \bibinfo{author}{\bibfnamefont{T.}~\bibnamefont{McJunkin}},
  \bibinfo{author}{\bibfnamefont{S.~F.} \bibnamefont{Neyens}},
  \bibinfo{author}{\bibfnamefont{E.~R.} \bibnamefont{MacQuarrie}},
  \bibnamefont{et~al.} (\bibinfo{year}{2020}),
  \bibinfo{note}{arXiv:2009.13572},
  \urlprefix\url{https://arxiv.org/abs/2009.04268}.

\bibitem[{\citenamefont{Borjans et~al.}(2021)\citenamefont{Borjans, Zhang, Mi,
  Cheng, Yao, Jackson, Edge, and Petta}}]{BorjansPRXQ2021}
\bibinfo{author}{\bibfnamefont{F.}~\bibnamefont{Borjans}},
  \bibinfo{author}{\bibfnamefont{X.}~\bibnamefont{Zhang}},
  \bibinfo{author}{\bibfnamefont{X.}~\bibnamefont{Mi}},
  \bibinfo{author}{\bibfnamefont{G.}~\bibnamefont{Cheng}},
  \bibinfo{author}{\bibfnamefont{N.}~\bibnamefont{Yao}},
  \bibinfo{author}{\bibfnamefont{C.}~\bibnamefont{Jackson}},
  \bibinfo{author}{\bibfnamefont{L.}~\bibnamefont{Edge}}, \bibnamefont{and}
  \bibinfo{author}{\bibfnamefont{J.}~\bibnamefont{Petta}},
  \bibinfo{journal}{PRX Quantum} \textbf{\bibinfo{volume}{2}},
  \bibinfo{pages}{020309} (\bibinfo{year}{2021}),
  \urlprefix\url{https://link.aps.org/doi/10.1103/PRXQuantum.2.020309}.

\bibitem[{\citenamefont{Das~Sarma et~al.}(2011)\citenamefont{Das~Sarma, Wang,
  and Yang}}]{DasSarmaPRB2011}
\bibinfo{author}{\bibfnamefont{S.}~\bibnamefont{Das~Sarma}},
  \bibinfo{author}{\bibfnamefont{X.}~\bibnamefont{Wang}}, \bibnamefont{and}
  \bibinfo{author}{\bibfnamefont{S.}~\bibnamefont{Yang}},
  \bibinfo{journal}{Phys. Rev. B} \textbf{\bibinfo{volume}{83}},
  \bibinfo{pages}{235314} (\bibinfo{year}{2011}),
  \urlprefix\url{https://link.aps.org/doi/10.1103/PhysRevB.83.235314}.

\bibitem[{\citenamefont{Yang et~al.}(2011)\citenamefont{Yang, Wang, and
  Das~Sarma}}]{YangPRB2011}
\bibinfo{author}{\bibfnamefont{S.}~\bibnamefont{Yang}},
  \bibinfo{author}{\bibfnamefont{X.}~\bibnamefont{Wang}}, \bibnamefont{and}
  \bibinfo{author}{\bibfnamefont{S.}~\bibnamefont{Das~Sarma}},
  \bibinfo{journal}{Phys. Rev. B} \textbf{\bibinfo{volume}{83}},
  \bibinfo{pages}{161301} (\bibinfo{year}{2011}),
  \urlprefix\url{https://link.aps.org/doi/10.1103/PhysRevB.83.161301}.

\bibitem[{\citenamefont{Stafford and Das~Sarma}(1994)}]{StaffordPRL1994}
\bibinfo{author}{\bibfnamefont{C.~A.} \bibnamefont{Stafford}} \bibnamefont{and}
  \bibinfo{author}{\bibfnamefont{S.}~\bibnamefont{Das~Sarma}},
  \bibinfo{journal}{Phys. Rev. Lett.} \textbf{\bibinfo{volume}{72}},
  \bibinfo{pages}{3590} (\bibinfo{year}{1994}),
  \urlprefix\url{https://link.aps.org/doi/10.1103/PhysRevLett.72.3590}.

\bibitem[{\citenamefont{DiVincenzo et~al.}(2000)\citenamefont{DiVincenzo,
  Bacon, Kempe, Burkard, and Whaley}}]{DiVincenzoNAT2000}
\bibinfo{author}{\bibfnamefont{D.~P.} \bibnamefont{DiVincenzo}},
  \bibinfo{author}{\bibfnamefont{D.}~\bibnamefont{Bacon}},
  \bibinfo{author}{\bibfnamefont{J.}~\bibnamefont{Kempe}},
  \bibinfo{author}{\bibfnamefont{G.}~\bibnamefont{Burkard}}, \bibnamefont{and}
  \bibinfo{author}{\bibfnamefont{K.~B.} \bibnamefont{Whaley}},
  \bibinfo{journal}{Nature} \textbf{\bibinfo{volume}{408}},
  \bibinfo{pages}{339} (\bibinfo{year}{2000}), ISSN \bibinfo{issn}{1476-4687},
  \urlprefix\url{https://doi.org/10.1038/35042541}.

\end{thebibliography}

%
%

\end{document}